\UseRawInputEncoding
\documentclass[amsmath,amssymb,10pt,aps,twocolumn]{revtex4-2} 
\usepackage[colorlinks=true]{hyperref}
 \usepackage{lipsum}
 \usepackage{graphicx}
 \usepackage{textcomp}
\usepackage[dvipsnames]{xcolor}
\usepackage{latexsym}
\usepackage{amsmath}
\usepackage{amsthm}
\usepackage{amssymb}
\usepackage{epstopdf}
\usepackage{enumitem}
\usepackage{setspace} 
\usepackage{dcolumn} 
\usepackage{bm} 
\usepackage{setspace} 
\usepackage{slashed}
\usepackage{color}
\usepackage{xcolor}
\usepackage{youngtab}
\usepackage{tikz}
\usepackage{tikz-3dplot}
\usepackage{braket}
\usepackage[version=4]{mhchem}
\usepackage{textcomp}
\usepackage[export]{adjustbox}

\usetikzlibrary{shapes,snakes,arrows,chains,matrix,positioning,scopes,calc}
\usetikzlibrary{decorations.markings}

\usepackage[tabbotcap]{subfigure}

\allowdisplaybreaks
 
\begin{document}

\title{Electric field control of a quantum spin liquid in weak Mott insulators
}

\author{Daniel J. Schultz}
\affiliation{Department of Physics, University of Toronto, Toronto, Ontario M5S 1A7, Canada}

\author{Alexandre Khoury}
\affiliation{Department of Physics, University of Toronto, Toronto, Ontario M5S 1A7, Canada}

\author{F\'elix Desrochers}
\affiliation{Department of Physics, University of Toronto, Toronto, Ontario M5S 1A7, Canada}

\author{Omid Tavakol}
\affiliation{Department of Physics, University of Toronto, Toronto, Ontario M5S 1A7, Canada}

\author{Emily Z. Zhang}
\affiliation{Department of Physics, University of Toronto, Toronto, Ontario M5S 1A7, Canada}

\author{Yong Baek Kim}
\affiliation{Department of Physics, University of Toronto, Toronto, Ontario M5S 1A7, Canada}

\date{\today}   

\begin{abstract}
The triangular lattice Hubbard model at strong coupling, whose effective spin model contains both Heisenberg and ring exchange interactions, exhibits a rich phase diagram as the ratio of the hopping $t$ to onsite Coulomb repulsion $U$ is tuned. This includes a chiral spin liquid (CSL) phase. Nevertheless, this exotic phase remains challenging to realize experimentally because a given material has a fixed value of $t/U$ which is difficult to tune with external stimuli. One approach to address this problem is applying a DC electric field, which renormalizes the exchange interactions as electrons undergo virtual hopping processes; in addition to creating virtual doubly occupied sites, electrons must overcome electric potential energy differences. Performing a small $t/U$ expansion to fourth order, we derive the ring exchange model in the presence of an electric field and find that it not only introduces spatial anisotropy but also tends to enhance the ring exchange term compared to the dominant nearest-neighbor Heisenberg interaction. Thus, increasing the electric field serves as a way to increase the importance of the ring exchange at constant $t/U$. Through density matrix renormalization group calculations, we compute the ground state phase diagram of the ring exchange model for two different electric field directions. In both cases, we find that the electric field shifts the phase boundary of the CSL towards a smaller ratio of $t/U$. Therefore, the electric field can drive a magnetically ordered state into the CSL. This explicit demonstration opens the door to tuning other quantum spin systems into spin liquid phases via the application of an electric field.
\end{abstract}

\maketitle

\section{Introduction}

A quantum spin liquid (QSL) is a ground state of a magnetically frustrated quantum spin system, where competing interactions obstruct long-range order and instead yield a state characterized by long-range entanglement, fractionalized excitations, and emergent gauge fields~\cite{balents_spin_2010, savary_quantum_2017, knolle_field_2019, zhou_quantum_2017, broholm_quantum_2020, wen_quantum_2002, wen_colloquium_2017, wen_choreographed_2019}. However, experimentally realizing a QSL is an exceedingly difficult task. Half a century after their initial proposal~\cite{anderson_resonating_1973}, experimentalists continue the quest for an indisputable experimental realization of a QSL. Because QSLs are typically very sensitive to the microscopic parameters of the system, it is exceedingly hard to find a material that exactly lies within a QSL regime. This motivates the search for external tuning parameters that may bring a quantum spin system from an ordered state into a QSL. Typically, this is attempted via the application of a magnetic field, external pressure, chemical doping, or even a periodic driver~\cite{gordon_theory_2019, sorensen_heart_2021, kim_revealing_2016, zheng_gapless_2017, baek_evidence_2017, kasahara_majorana_2018, lee_magnetic_2020, chern_magnetic_2020, li_magnetic_2022, liu_pseudospin_2018, sano_kitaev-heisenberg_2018, rayyan_field-induced_2023, sur_driven_2022, kuhlenkamp_tunable_2022, claassen_dynamical_2017, wang_producing_2017, quito_floquet_2021, quito_polarization_2021, kobayashi_light-induced_2021}. However, one stimulus which has received relatively little attention is that of an electric field.

One relevant context in which to study the possibility of stabilizing QSLs via application of an electric field is weak Mott insulators on the triangular lattice. Deep in the Mott insulating phase, the system may be described by an effective Heisenberg Hamiltonian. Although frustrated, the ground state is given by a spiral 120{\textdegree} order. Closer to the Mott transition, further neighbor Heisenberg couplings and ring exchange interactions arise due to increased charge fluctuations~\cite{macdonald_t_1988, takahashi_half-filled_1977, delannoy_neorder_2005}. These ring exchange interactions lead to enhanced frustration, which may be sufficient to melt the long-range 120{\textdegree} order into a QSL. Indeed, recent density matrix renormalization group (DMRG) calculations have shown that the system stabilizes a chiral spin liquid (CSL) at physically relevant parameters~\cite{cookmeyer_four-spin_2021, szasz_chiral_2020, szasz_phase_2021}. CSLs are a well-studied class of QSLs that spontaneously break time-reversal symmetry and support a gapless chiral edge state~\cite{bose_chiral_2023, shirakawa_ground-state_2017, zhang_bosonic_2021,gong_emergent_2014, messio_kagome_2012, wietek_chiral_2017, tang_spectra_2022, tocchio_hubbard_2021, chen_quantum_2022, kadow_hole_2022, zhou_quantum_2022, willsher_magnetic_2023, hickey_competing_2015, hickey_haldane-hubbard_2016, hickey_emergence_2017, sahebsara_hubbard_2008, momoi_possible_1997, bauer_chiral_2014, gong_chiral_2019}.

Although an external magnetic field may appear to be a more natural tuning parameter to stabilize the CSL because it explicitly introduces a chiral exchange term in the effective spin model~\cite{huang_magnetic_2023}, this may also polarize the system due to the relatively large energy scale of the Zeeman term~\cite{sen_large-_1995}. In contrast, an electric field couples only to the electron's charge. The electric field modifies the energy cost incurred by virtual hopping processes in the Mott insulating phase as the electron moves with or against the field. Indeed, a DC electric field has previously been shown to modify the nearest-neighbor Heisenberg exchange parameters~\cite{takasan_control_2019, furuya_control_2021, furuya_electric-field_2021}. However, understanding the impact of an electric field on four-spin terms, which are responsible for the chiral spin liquid, is not yet explored.

This work considers the single-band Hubbard model on the triangular lattice at half-filling in a spatially uniform DC electric field. Taking the strong coupling limit, we derive a spin model of nearest, next nearest, and third nearest-neighbor Heisenberg couplings and ring exchange interactions. These interactions, which are typically the same along the different bonds, become anisotropic in real space. We also find that the ring exchange interactions are generally enhanced relative to the nearest-neighbor Heisenberg couplings as the electric field increases. This provides a mechanism to enhance ring exchange interactions relative to Heisenberg couplings without changing $t/U$. Since nearest-neighbor Heisenberg coupling favors 120{\textdegree} magnetic order~\cite{momoi_magnetization_1999, kubo_ground_1997} and ring exchange interactions stabilize a CSL~\cite{cookmeyer_four-spin_2021}, the electric field serves as an effective way to tune towards the CSL. Using DMRG calculations, we derive the zero-temperature phase diagram of the model as a function of $t/U$ and electric field. Most notably, we demonstrate that the regime of the chiral spin liquid is modified as the electric field is tuned, meaning that an initially magnetically ordered state may enter the chiral spin liquid at fixed $t/U$. Hence, the application of an electric field in a weak Mott insulator offers a novel avenue to reach QSL.

The remainder of the paper is organized as follows. Section II introduces the Hubbard model in an electric field and derives the effective Heisenberg and ring exchange spin model in the strong coupling limit. Then, in Section III, we establish the quantum phase diagram without an electric field. In Section IV, we introduce the electric field in two different directions. We conclude in Section V with a discussion of our results.

\section{Effective Spin Model}

\subsection{Hubbard Model in an Electric Field}

We start by considering the single-band Hubbard model at half-filling in a static and spatially uniform electric field~\cite{takasan_control_2019}. In this general setting, this discussion applies equally well to any lattice. This model is given by
\begin{equation}
H = -\sum_{i,j;\sigma}t_{ij} c^\dagger_{i\sigma} c_{i\sigma} + U\sum_i n_{i\uparrow} n_{i\downarrow} + \sum_{i\sigma} \Phi_i n_{i\sigma} \label{eq:hubbard_model}
\end{equation}
where $\sigma = \uparrow,\downarrow$ represents the spin of the electrons, $i,j$ run over all sites in the triangular lattice, and $n_{i\sigma} = c^\dagger_{i\sigma} c_{i\sigma}$ is the number operator for electrons on site $i$ with spin $\sigma$. $U$ is the on-site Coulomb repulsion, $\Phi_i$ is the electric potential energy at site $i$, and the hopping matrix is given by $t_{ij}$, which we will later set to 
\begin{equation}
t_{ij} = \begin{cases} t, & i,j \, \text{nearest-neighbors} \\ 0, & \text{else} \end{cases}.
\end{equation}
In the limit $t/U \ll 1$ and also $|\Delta\Phi|/U \ll 1$, where $\Delta\Phi$ is the electric potential energy difference between neighboring sites, we can perform a Schrieffer-Wolff transformation to find an effective Hamiltonian governing the low-energy behavior~\cite{takasan_control_2019}. The virtual high-energy fluctuations are induced by the portion of the Hamiltonian that can change the number of doubly occupied sites. We split the kinetic term into three parts $T^+,T^-,T^0$. It can be identified that $T^+$ increases the number of doubly occupied sites, $T^-$ decreases the number of doubly occupied sites, and $T^0$ does not change the number of doubly occupied sites:
\begin{align}
T^+ ={}& \sum_{ij} T^+_{ij},\quad T^- = \sum_{ij} T^-_{ij}, \quad T^0 = \sum_{ij} T^0_{ij}, \\
T^+_{ij} ={}& -\sum_\sigma t_{ij} n_{i,-\sigma}c^\dagger_{i\sigma}c_{j\sigma}h_{j,-\sigma}, \\
T^-_{ij} ={}& -\sum_\sigma t_{ij} h_{i,-\sigma} c^\dagger_{i\sigma}c_{j\sigma} n_{j,-\sigma}, \\
T^0_{ij} ={}& -\sum_{\sigma}t_{ij} \left[h_{i,-\sigma}c^\dagger_{i\sigma}c_{j\sigma}h_{j,-\sigma} + n_{i,-\sigma}c^\dagger_{i\sigma}c_{j\sigma}n_{j,-\sigma}\right],
\end{align}
where $h_{i\sigma} = 1-n_{i\sigma}$ is the hole number operator. We set the perturbation, which produces the charge fluctuations, to be $W = T^+ + T^-$ and the base Hamiltonian as $H_0 = T_0 + H_U + H_E$, whereby the potential energies are

\begin{equation}
H_U = U\sum_i n_{i\uparrow} n_{i\downarrow}, \quad H_E = \sum_{i\sigma} \Phi_i n_{i\sigma}.
\end{equation}
These are included within the base Hamiltonian because they do not cause charge fluctuations.

\subsection{Canonical Transformation}

In order to find a low-energy description of the theory, we perform a canonical transformation of our Hamiltonian. The transformation is defined by a generator $S$ as
\begin{equation}
H_\text{eff} = e^{iS}H e^{-iS}. \label{eq:canonical_transformation}
\end{equation}
\noindent Here, $S$ may be computed order-by-order in $t/U$, and $H_\text{eff}$ is the new effective Hamiltonian whose fluctuations to the high-energy sector with doubly occupied sites are exactly eliminated to order $O(t^3/U^2)$. The expansion of the generator $S$ is given by
\begin{equation}
S = S^{(1)} + S^{(2)} + S^{(3)} + \cdots
\end{equation}
\noindent where $S^{(n)} \propto (t/U)^n$. The detailed procedure for calculating the generator of the canonical transformation is in Appendix~\ref{app:sc_expansion}. We compute $S$ up to third order in $t/U$. Expanding the effective Hamiltonian in Eq.~\eqref{eq:canonical_transformation} yields 
\begin{equation}
H_\text{eff} = H + [iS,H] + \frac{1}{2}[iS,[iS,H]] + \cdots \label{eq:Heff_expansion}
\end{equation}
We solve for $S^{(n)}$ such that no terms in the effective Hamiltonian take us out of the singly occupied sector. Such terms have an unequal number of $T^+$ and $T^-$ operators. For example, at first order we pick $T^++T^- + [iS^{(1)},H_U+H_E] = 0$ to cancel the charge fluctuations. The solution for the generator of the canonical transformation generalizes the typical expressions without an electric field by introducing factors that encode the energy difference obtained when an electron changes its potential energy by moving with or against the electric field. The expressions for the generators are 
\begin{align}
iS^{(1)} ={}& \frac{1}{U}\sum_{ij} \Lambda_{ij}(T^+_{ij} - T^-_{ji}), \label{eq:S1} \\
iS^{(2)} ={}& \frac{1}{U^2}\sum_{ijpq} \Omega_{ijpq}\left([T^+_{ij},T^0_{pq}] + [T^-_{ji},T^0_{qp}]\right), \label{eq:S2}
\end{align}
\begin{widetext}
\begin{align}
iS^{(3)} ={}& \frac{1}{U^3}\sum_{ijpqab} \Xi_{ijpqab} ([[T^+_{ij},T^0_{pq}],T^0_{ab}] - [[T^-_{ji},T^0_{qp}],T^0_{ba}]) + \frac{1}{2U^3}\sum_{ijpqab}\Xi'_{ijpqab}([[T^+_{ij},T^0_{pq}],T^+_{ab}] - [[T^-_{ji},T^0_{qp}],T^-_{ba}]) \nonumber \\
&+ \frac{1}{3U^3}\sum_{ijpqab}\Lambda_{ij}(\Lambda_{pq} + \Lambda_{ab})\Lambda_{ijpqba}([T^+_{ij},[T^+_{pq},T^-_{ba}]] + [T^-_{ji},[T^+_{ab},T^-_{qp}]]). \label{eq:S3}
\end{align}
\end{widetext}
The information about the electric field is contained within the various factors of the form
\begin{align}
\Lambda_{ij} ={}& \frac{1}{1+\Phi_{ij}},\\
\Omega_{ijpq} ={}& \frac{\Lambda_{ij}}{1+\Phi_{ij}/U + \Phi_{pq}/U},\\
\Xi_{ijpqab} ={}& \frac{\Omega_{ijpq}}{1+\Phi_{ij}/U + \Phi_{pq}/U + \Phi_{ab}/U},\\
\Xi'_{ijpqab} ={}& \frac{\Omega_{ijpq}}{2 + \Phi_{ij}/U + \Phi_{pq}/U + \Phi_{ab}/U},\\ \Lambda_{ijpqab} ={}& \frac{1}{1 + \Phi_{ij}/U + \Phi_{pq}/U + \Phi_{ab}/U}.
\end{align}
%

\subsection{Effective Hamiltonian}

\begin{figure}
\centering
\includegraphics[scale = 1]{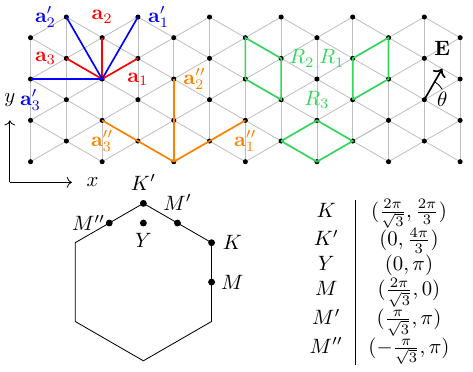}
\caption{Top: The triangular lattice, its primitive lattice vectors $\mathbf{a}_1 = (\frac{\sqrt{3}}{2},\frac{1}{2})$ and $\mathbf{a}_2 = (0,1)$ and our convention for labeling of the bonds and rings. Bottom: Brillouin zone and high symmetry points of the triangular lattice. \label{fig:lattice_and_BZ}}
\end{figure}

The effective Hamiltonian obtained, which is obtained by replacing Eqs.~\eqref{eq:S1}-\eqref{eq:S3} into Eq.~\eqref{eq:Heff_expansion}, truncating to order $O(t^4/U^3)$, and then projecting to the singly occupied subspace, yields a spin model. We need to pick a specific form of the electric potential to write down this spin model in a compact form. We choose the potential $\Phi_i$ derived from the uniform electric field $\mathbf{E} = E(\cos(\theta+\pi/6),\sin(\theta+\pi/6))$, where $\theta$ is measured with respect to $\mathbf{a}_1 = (\sqrt{3}/2,1/2)$ (see Fig.~\ref{fig:lattice_and_BZ} for our coordinate convention). The resulting effective spin model is 
\begin{align}
H_\text{eff} ={}& \sum_{n=1}^3 J^{(n)}_1(\theta)\sum_{i} \textbf{S}_i\cdot\textbf{S}_{i+\mathbf{a}_n} \label{eq:spin_model} \\
&+ \sum_{n=1}^3 J^{(n)}_2(\theta)\sum_{i} \textbf{S}_i\cdot\textbf{S}_{i+\mathbf{a}_n'} \nonumber \\
&+\sum_{n=1}^3 J^{(n)}_3(\theta)\sum_{i} \textbf{S}_i\cdot\textbf{S}_{i+\mathbf{a}_n''} \nonumber \\
&+ H_\text{ring}, \nonumber
\end{align}
where the sum over $n$ represents the sum over inequivalent bonds in the presence of an electric field, and the couplings $J_1,J_2,J_3$ are the first, second, and third nearest neighbor Heisenberg interactions (see Fig.~\ref{fig:lattice_and_BZ}). The particular forms of the couplings and their dependence on the electric field are lengthy and listed in Appendix~\ref{app:couplings}. $H_\text{ring}$ is the ring exchange term generated by the $t/U$ expansion:
\begin{align}
H_\text{ring} ={}& \sum_{n=1}^3 J^{(n)}_r(\theta) \sum_{i,j,k,\ell\in R}[(\textbf{S}_i\cdot\textbf{S}_j)(\textbf{S}_k\cdot\textbf{S}_\ell) \notag\\
&+ (\textbf{S}_i\cdot\textbf{S}_\ell)(\textbf{S}_j\cdot\textbf{S}_k) - (\textbf{S}_i\cdot\textbf{S}_k)(\textbf{S}_j\cdot\textbf{S}_\ell)],
\end{align}
where the detailed expressions of $J^{(n)}_r(\theta)$ are shown in Appendix~\ref{app:couplings}. We note that the exchange constants satisfy the typical values for the triangular lattice in the limit $E\to 0$ \cite{cookmeyer_four-spin_2021}
\begin{equation} \label{eq:couplings_e0}
J_1 = \frac{4t^2}{U} - \frac{28t^4}{U^3},\quad J_2 = J_3 = \frac{4t^4}{U^3},\quad J_r = \frac{80t^4}{U^3}.
\end{equation}

\begin{figure}
\centering
\includegraphics[scale=0.26]{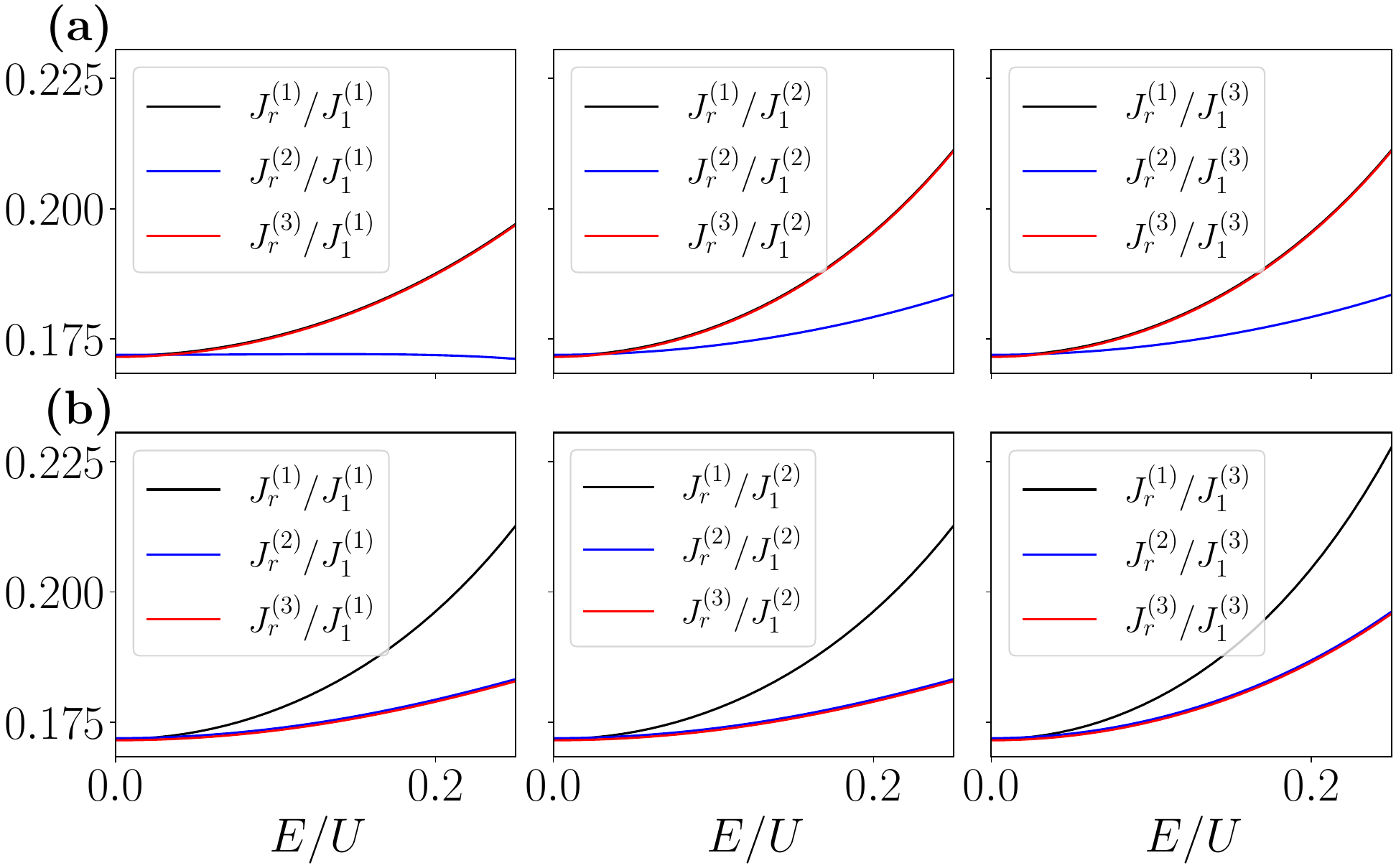}
\caption{ Electric field dependence for the ratio of the ring exchange coupling $J_{r}^{(m)}$ on ring $R_m$ to the nearest-neighbor Heisenberg coupling $J_{1}^{(n)}$ along the $\mathbf{a}_n$ bond direction with an electric field (a) along bond $\mathbf{a}_1$ ($\theta = 0$) and (b) halfway between bond $\mathbf{a}_1$ and $\mathbf{a}_2$ ($\theta=\pi/6$). We slightly shift overlapping curves to make them visible. In actuality, some of the curves coincide due to symmetry considerations.}
\label{fig:coupling_e_dependence}
\end{figure}

The most important feature of the model is that, although the electric field amplifies many couplings, the ring exchange grows faster than the other ones as $E/U$ increases. This behavior is illustrated in Fig.~\ref{fig:coupling_e_dependence}. It can be noticed that for the case of the electric field along $\mathbf{a}_1$ ($\theta = 0$), almost all of the $J_r/J_1$ ratios increase except for one of them. Similarly for $\theta = \pi/6$, every $J_r/J_1$ ratio increases. This offers a way to tune the ratio $J_r/J_1$ for most directions, which may potentially send the system into the CSL. 

Some intuition about the electric field dependence of the ring exchange strength can be gained by making the following observation. The electric field only enhances a coupling strength if the electron's potential energy changes along the direction of a virtual hopping. For example, the enhancement of the nearest-neighbor Heisenberg interaction is the strongest if the electric field is parallel to the bond but is not modified at all if it is perpendicular. Although the ring exchange couplings are very complicated, a similar idea emerges. If the electric field is parallel to $\mathbf{a}_1$ ($\theta=0$), then the ring exchange interactions for the rings $R_1$ and $R_3$ (see Fig.~\ref{fig:lattice_and_BZ}) are equally enhanced because they both contain $\mathbf{a}_1$ bonds, but $R_2$ contains only $\mathbf{a}_2$ and $\mathbf{a}_3$ bonds, so it is not as strongly affected. A similar scenario occurs if the electric field is applied halfway between the bonds, say $\theta=\pi/6$. In this case, both $R_2$ and $R_3$ have bonds perpendicular to the field direction, but the electric field has a non-zero projection on all four bonds in $R_1$. For this reason, the ring exchange for $R_1$ is enhanced the most.

\section{Quantum Phase Diagram for $E/U = 0$} \label{sec:zero_efield}

We use DMRG to determine the ground state of the effective spin model using the TenPy package~\cite{hauschild_efficient_2018}.  In two dimensions, DMRG uses the one-dimensional matrix product state (MPS) representation by  snaking the MPS through a two dimensional unit cell of sites. We use iDMRG, wherein the MPS unit cell has a length of $L_x=2$ sites in the $\mathbf{a}_1$ direction and length $L_y = 6$ sites in the $\mathbf{a}_2$ direction, but the unit cell is repeated infinitely along $\mathbf{a}_1$. This facilitates a study of our spin system on a cylinder with circumference $L_y$, and allows us to compute long-ranged correlations along the $\mathbf{a}_1$ direction. Our model has both full $SU(2)$ rotational symmetry, but only a $U(1)$ symmetry subgroup of rotation is explicitly encoded in constructing the MPS. Accordingly, we have $S^z$ conservation, and all results lie in the $S^z_{\text{tot}} = 0$ sector. The simulations were done with bond dimension $b = 1600$.

In our simulations, we further introduce a chiral symmetry-breaking term of the form 
\begin{align}
H_\chi = J_\chi \sum_{i,j,k\in \triangleright,\triangleleft} \mathbf{S}_i\cdot(\mathbf{S}_j\times\mathbf{S}_k).
\end{align}
Here, $i,j,k$ are three sites arranged counterclockise around the two triangle types $\triangleright$ and $\triangleleft$. $H_\chi$ is not a part of the $t/U$ expansion (unless there is an external magnetic field) and is introduced to gently break chiral symmetry to allow the possibility of finding the chiral spin liquid. The coupling $J_\chi$ is set to a very small number in the initialization of the simulations and is subsequently set to zero. Further details regarding the DMRG are given in Appendix~\ref{app:dmrg}.

\begin{figure}
    \centering
    \includegraphics[scale=0.55]{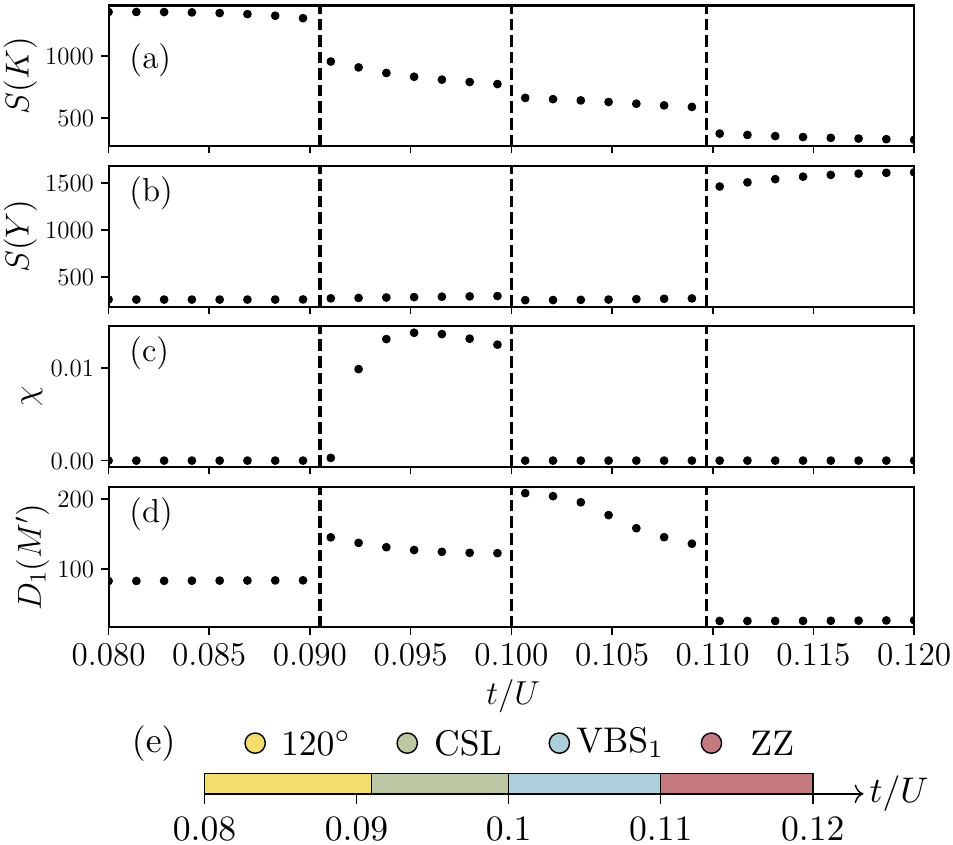}
    \caption{Phase diagram and values of correlators as a function of $t/U$ for zero electric field. (a) Spin structure factor at the $K$ point in the Brillouin zone (BZ), (b) Spin structure factor at the $Y$ point in the BZ, (c) Scalar chirality, (d) Dimer structure factor along the $\mathbf{a}_2$ bond at the $M''$ point in the BZ, (e) Phase diagram as a function of $t/U$.} \label{fig:E0phase}
\end{figure}

We first establish the results of the simulations in the absence of the electric field. Previously, a phase diagram has been established for the $J_1$-$J_r$ model~\cite{cookmeyer_four-spin_2021}. A sequence of transitions from spiral order to a chiral spin liquid (CSL), then to a valence bond solid (VBS), and ultimately to a zigzag ordered phase has been predicted as the ratio $J_r/J_1$ increases. The ring exchange model obtained from the $t/U$ expansion is expected to be similar because $J_2$ and $J_3$ are very small compared to $J_r$ (i.e., about $J_r/20$), and the ratio $J_r/J_1$ increases monotonically as a function of $t/U$. Indeed, we find a similar phase diagram to that by Cookmeyer et al.\cite{cookmeyer_four-spin_2021}, as shown in Fig.~\ref{fig:E0phase}. At small $t/U$, we find a 120{\textdegree} ordered phase. As we increase $t/U$, we enter a chiral spin liquid phase, then a valence bond solid phase. Lastly, we enter a zigzag-ordered phase at large $t/U$. As illustrated in Fig.~\ref{fig:E0phase}(a)-(d), the phase transitions can be clearly identified by jumps in the spin structure factor and dimer structure factors at specific high symmetry points, as well as a non-zero average scalar chirality indicating the CSL. We expand further on the four different phases at zero electric fields and how they can be identified.

\textit{Spiral order:} The 120{\textdegree} ordered phase is a classical long-range ordered phase. It can be identified by strong peaks in the spin structure factor 
\begin{equation}
S(\bm{k}) = \sum_{ij} e^{i\bm{k}\cdot(\mathbf{R}_i - \mathbf{R}_j)} (\langle \mathbf{S}_i \cdot \mathbf{S}_j \rangle - \langle \mathbf{S}_i \rangle \langle \mathbf{S}_j \rangle),
\end{equation}
at $K,K'$ points in the Brillouin zone (i.e., the ordering wavevector). The classical spin configuration is a three-site unit cell with the three spins angled at 120{\textdegree} relative to one another. When the electric field is non-zero, the ordering wavevector shifts slightly from the high symmetry $K, K'$ points and changes continuously with the electric field strength. To be general, we shall accordingly refer to this phase as a spiral order. The spiral phase should be understood as a classically ordered magnet with an ordering wavevector close to the $K$ point adiabatically connected to the 120{\textdegree} order. The 120{\textdegree} order name will only be used when the spiral phase is at zero electric field and has a commensurate order. 

\begin{figure}
    \centering
    \includegraphics[scale=0.48]{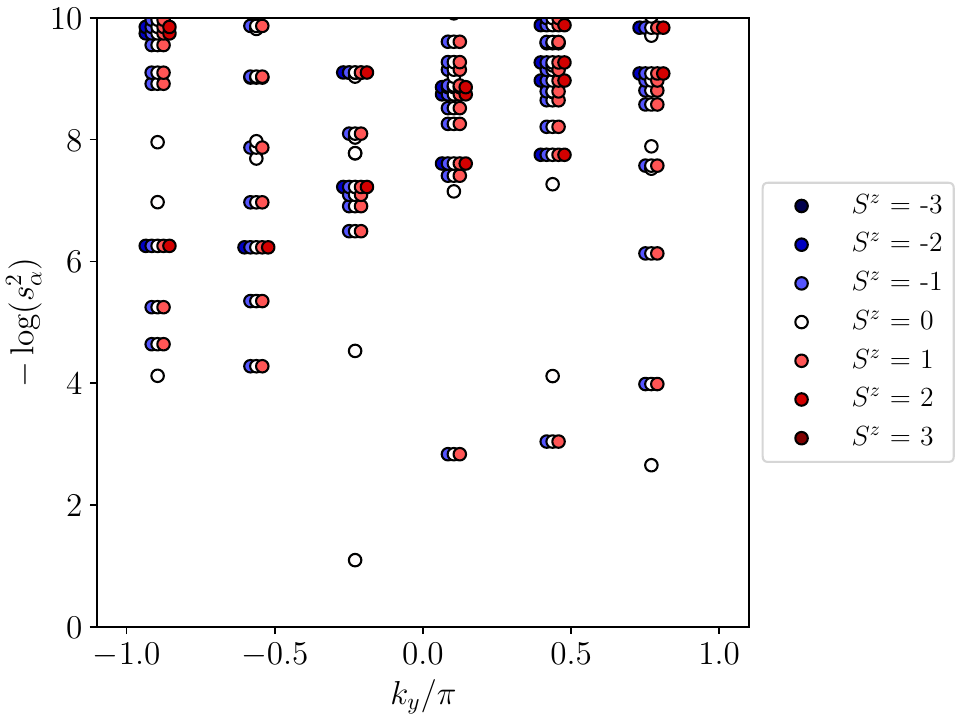}
    \caption{The momentum resolved entanglement spectrum of the chiral spin liquid at $t/U = 0.097$ and $E/U = 0$. The $y$ axis is $- \log(s_\alpha^2)$, where $s_\alpha$ are the Schmidt values, and the $x$-axis is the momentum $k_y$ around the cylinder.} 
    \label{fig:CSL_ES}
\end{figure}

\textit{Chiral spin liquid:} The CSL is a QSL that supports deconfined fractional excitations and breaks time-reversal symmetry. It can identified by the non-zero scalar chirality $\chi$, defined by
\begin{equation}
\chi = \frac{1}{2L_xL_y}\sum_{i,j,k\in \triangleright,\triangleleft} \langle \mathbf{S}_i \cdot (\mathbf{S}_j \times \mathbf{S}_k) \rangle,
\end{equation}
where we average over all triangles of the lattice (there are $2L_xL_y$ many triangles). The non-zero spin chirality is not definitive evidence for the CSL since classical non-coplanar orders also have $\chi\ne 0$. More conclusive evidence for the CSL can be obtained by investigating the entanglement spectrum of this state shown in Fig.~\ref{fig:CSL_ES}. The spectrum breaks inversion symmetry, which is indicative of the presence of the chiral edge modes. It is identical to the one reported in Ref.~\cite{cookmeyer_four-spin_2021} where it has been argued that each of the levels with spin quantum numbers $\left|S^z\right| \in\{0,1,2\}$ shows the degeneracy pattern expected for the Kalmeyer-Laughlin wavefunction~\cite{Li_entanglement_2008}.

\textit{Valence bond solid 1:} A valence bond solid describes a covering of the lattice with singlet states of two spins. The triangular lattice has six simple nearest-neighbor dimer coverings, depicted pictorially in Appendix~\ref{app:dimer_coverings}. To probe different dimer coverings, we can compute different dimer structure factors defined in terms of dimer operators $D^n_i = \mathbf{S}_i \cdot \mathbf{S}_{i + \mathbf{a}_n}$ by
\begin{equation}
D_n(\bm{k}) = \sum_{ij} e^{i\bm{k}\cdot(\mathbf{R}_i - \mathbf{R}_j)}(\langle D^n_i D^n_j \rangle - \langle D^n_i \rangle \langle D^n_j \rangle).
\end{equation}
A VBS state has sharp peaks in its dimer structure factors. The dominant dimer covering correlation can further be identified by the position by the peak locations of three different dimer structure factors, as explained in Appendix~\ref{app:dimer_coverings}. For the valence bond solid 1 (VBS$_1$) state, the dominant dimer-dimer correlations have been determined to be along the $\mathbf{a}_1$ and $\mathbf{a}_3$ bonds (see Figs.~\ref{fig:dimers_a}-\ref{fig:dimers_c}). As a first-order approximation, one can then think schematically of VBS$_1$ as an equal-weight superposition of the two dominant dimer coverings
\begin{equation}
\ket{\Psi_{\text{VBS}_1}} \sim \includegraphics[scale=0.6,valign=c]{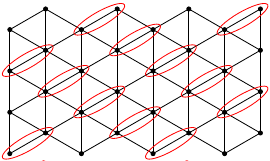}+\includegraphics[scale=0.6,valign=c]{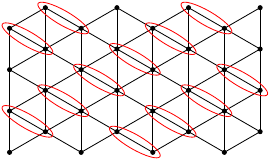}. \label{eq:VBS1_WF}
\end{equation}

\begin{figure}
    \centering
    \includegraphics{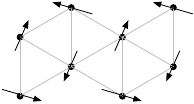}
    \caption{Zigzag spin configuration. The order has a 4 site unit cell, and two unit cells are shown.}
    \label{fig:zigzag_configuration}
\end{figure}

\textit{Zigzag:} The zigzag phase is a long-range ordered phase with ordering momentum at the $Y$ point in the Brillouin zone. We identify it by strong peaks in the static spin structure factor at this high symmetry point. The spin configuration of this order is given by Fig.~\ref{fig:zigzag_configuration}.

A detailed characterization of all phases identified in this work is provided in Appendix~\ref{app:dmrg_data}. There, we present the spin-spin correlations in real and momentum space, the three dimer-dimer correlations in real and momentum space, as well as the entanglement spectrum as a function of transverse momentum and total spin quantum number $S^{z}$ for all phases. 

\section{Quantum Phase Diagram for $E/U \neq 0$} \label{sec:nonzero_efield}

In the previous two sections, we made the following observations: the electric field introduces spatial anisotropy in the spin exchange Hamiltonian and (typically) increases the ring exchange relative to the Heisenberg exchange for specific directions depending on the electric field orientation. These effects lead to a competition between two opposite trends. On the one hand, the resulting spatial anisotropy reduces the amount of frustration and would favor magnetic order. On the other hand, increasing the magnitude of the ring exchange (and correspondingly increasing the frustration) can drive the system towards a  CSL. This naturally leads to the question: Is the increased frustration due to the enhanced ring exchange sufficient to overcome the reduced frustration caused by the spatial anisotropy and shift the CSL towards smaller $t/U$? That is the question that we will address in this section.

Having established the phase diagram without an electric field, we now set $E\neq 0$ and modify the effective exchange couplings. The angle $\theta$ of the electric field is defined with respect to the vector $\mathbf{a}_1 = (\sqrt{3},1)/2$, as shown in Fig.~\ref{fig:lattice_and_BZ}. We will consider two different high symmetry directions of the electric field, namely $\theta = 0$ (aligned along $\mathbf{a}_1$), and $\theta = \pi/6$ (aligned along $\mathbf{a}_1 + \mathbf{a}_2$). 

\subsection{Electric field along $\theta = 0$}

\begin{figure}
\centering
\includegraphics[scale=0.45]{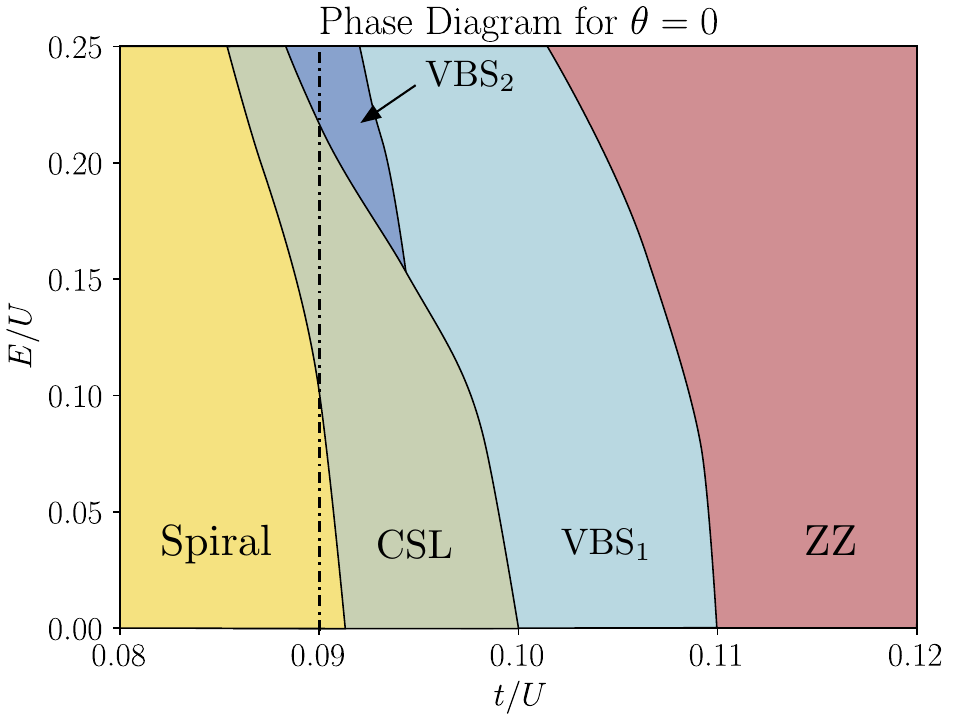}
\caption{Phase diagram with an electric field along the $\mathbf{a}_1$ bond ($\theta = 0$). Observables along the vertical dash-dotted line are depicted in Fig.~\ref{fig:t09_theta0_phase}}
\label{fig:phasediagram0}
\end{figure}

As illustrated in Fig.~\ref{fig:phasediagram0}, aligning the electric field along $\mathbf{a}_1$ makes the phase boundaries shift as the electric field increases. More precisely, the phase boundaries are generally shifted towards smaller $t/U$. The general shifting of the phase boundaries to smaller $t/U$ as the field is increased is logical, considering that the electric field enhances the relative importance of the ring exchange. The effect of a non-zero electric field is thus naively comparable to a direct increase of $t/U$. This general shifting of the phase boundaries implies that starting from a classically ordered state, one can promote the system to a CSL by applying an external electric field. Fig.~\ref{fig:phasediagram0} explicitly demonstrates that a DC electric field can be used as an efficient tuning parameter to promote QSLs. This possibility to stabilize QSL by applying an electric field is the central point we wish to convey in this work.

\begin{figure}
    \centering
    \includegraphics[scale=0.5]{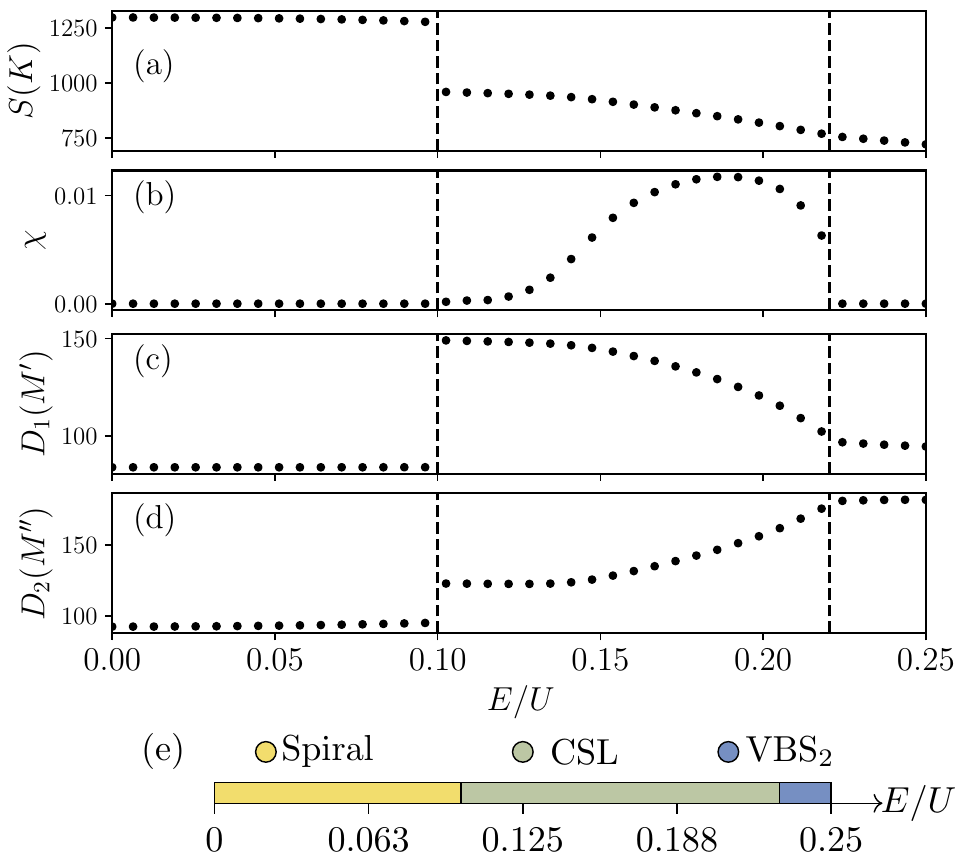}
    \caption{Phase diagram and values of correlators at fixed $t/U = 0.09$ as a function of $E/U$ for the electric field pointing along the $\mathbf{a}_1$ bond ($\theta = 0$). (a) The spin structure factor at the $K$ point in the BZ, (b) The scalar chirality. (c) The dimer structure factor along the $\mathbf{a}_1$ bond at the $M'$ point in the BZ. (d) The dimer structure factor along the $\mathbf{a}_2$ bond at the $M''$ point in the BZ. (e) The overall phase diagram for this line.}
    \label{fig:t09_theta0_phase}
\end{figure}

Let us see in more detail how the chiral spin liquid arises as we travel along a vertical line. The evolution of different observables as the electric field is increased for a fixed value of $t/U = 0.09$ is presented in Fig.~\ref{fig:t09_theta0_phase} (this corresponds to the dash-dotted vertical line in Fig.~\ref{fig:phasediagram0}). Starting from the spiral phase, it transitions to a CSL with non-zero chirality at around $E/U = 0.1$. However, because we are traveling roughly parallel to the phase boundary, the order parameter grows slowly until it reaches its typical value of around $\chi = 0.01$ at around $E/U = 0.18$. It has been explicitly determined that this CSL is the same for zero electric field by directly comparing their entanglement spectra. For any non-zero electric field, the CSL entanglement spectrum remains identical to Fig.~\ref{fig:CSL_ES}, and we do not see any sign of a phase transition. Finally, as we leave the CSL at around $E/U = 0.22$, we enter a new phase, labeled VBS$_2$, that does not occur at zero electric field.

\begin{figure*}
\includegraphics[width=1.00\textwidth]{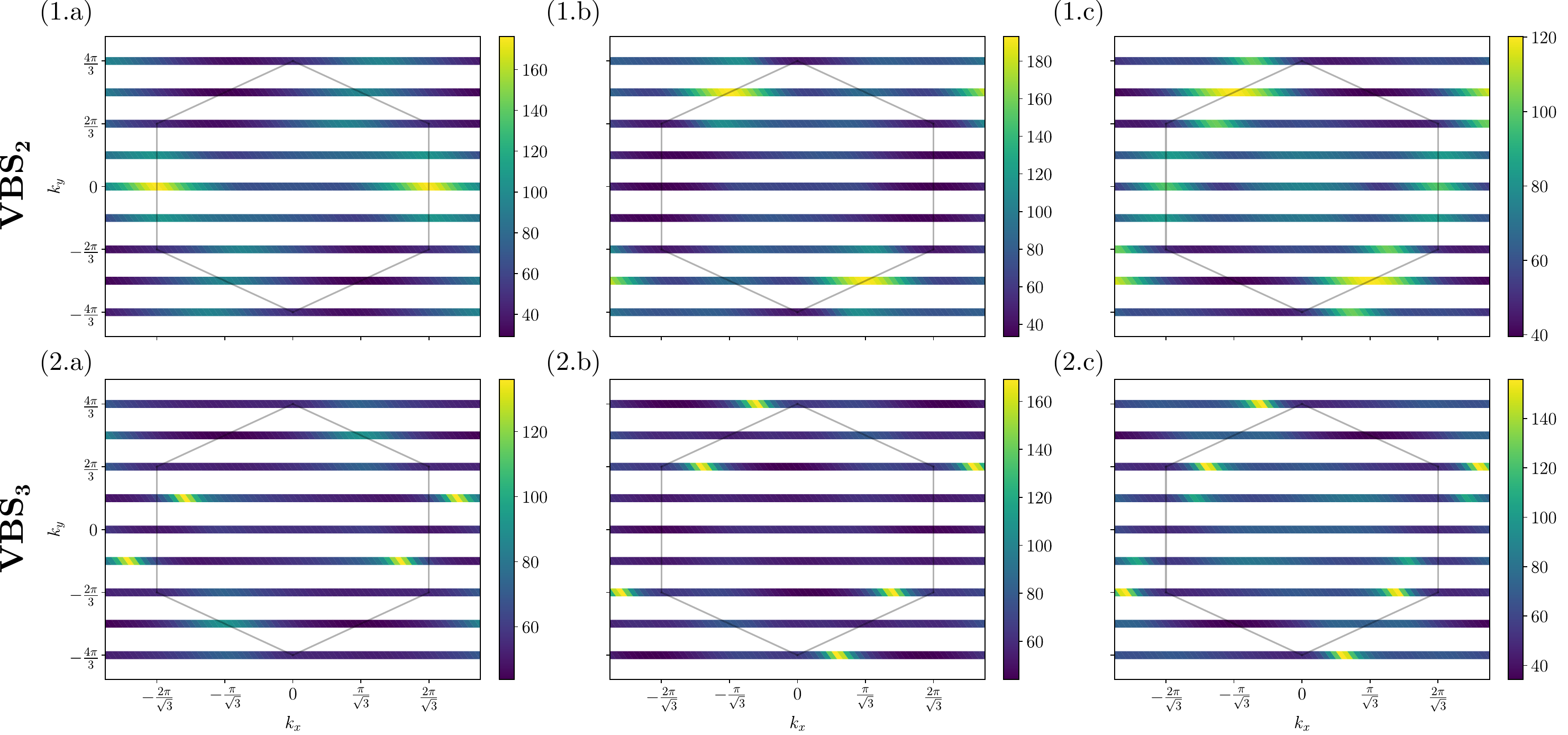}
\caption{Dimer structure factor (a) $D_1(\bm{k})$, (b) $D_1(\bm{k})$, and (c) $D_3(\bm{k})$ for (1) the VBS$_2$ phase with $t/U = 0.099$ and $E/U = 0.200$ at an angle  $\theta=0$ and (2) the VBS$_3$ phase with $t/U = 0.102$ and $E/U = 0.250$ at an angle $\theta=\pi/6 = 0.524$.} \label{fig:vbs2_vbs3}
\end{figure*}

The VBS$_2$ phase, just like the VBS$_1$ phase, is a VBS phase as seen from its strong peaks in the dimer structure factors at high symmetry points in the Brillouin zone (see panels (1.a)-(1.c) of Fig.~\ref{fig:vbs2_vbs3}). In order to determine the dominant dimer coverings in the wave function, we compare the peaks of the dimer structure factor with those expected for the various dimer coverings in (see Fig.~\ref{fig:dimers_a}-\ref{fig:dimers_c} of Appendix~\ref{app:dimer_coverings}). The highest peaks are in the dimer structure factor for singlets along the $\mathbf{a}_2$ direction. Thus, we can conclude that the translation vectors of the dimer covering are given by $\mathbf{a}_1$ and $2\mathbf{a}_2 - \mathbf{a}_1$. From this analysis, one can then consider that the VBS$_2$ wavefunction is conceptually of the form
\begin{equation}
\ket{\Psi_{\text{VBS}_2}} \sim \includegraphics[scale=0.6,valign=c]{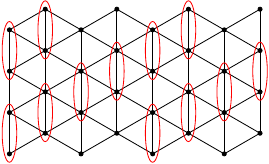}. \label{eq:VBS2_WF}
\end{equation}
Of course, there may be contributions from other dimer coverings, but we only include the dominant one to provide an intuitive understanding of the different VBS phases.
 

\subsection{Electric field along $\theta = \pi/6$}

\begin{figure}[b]
\centering
\includegraphics[scale=0.45]{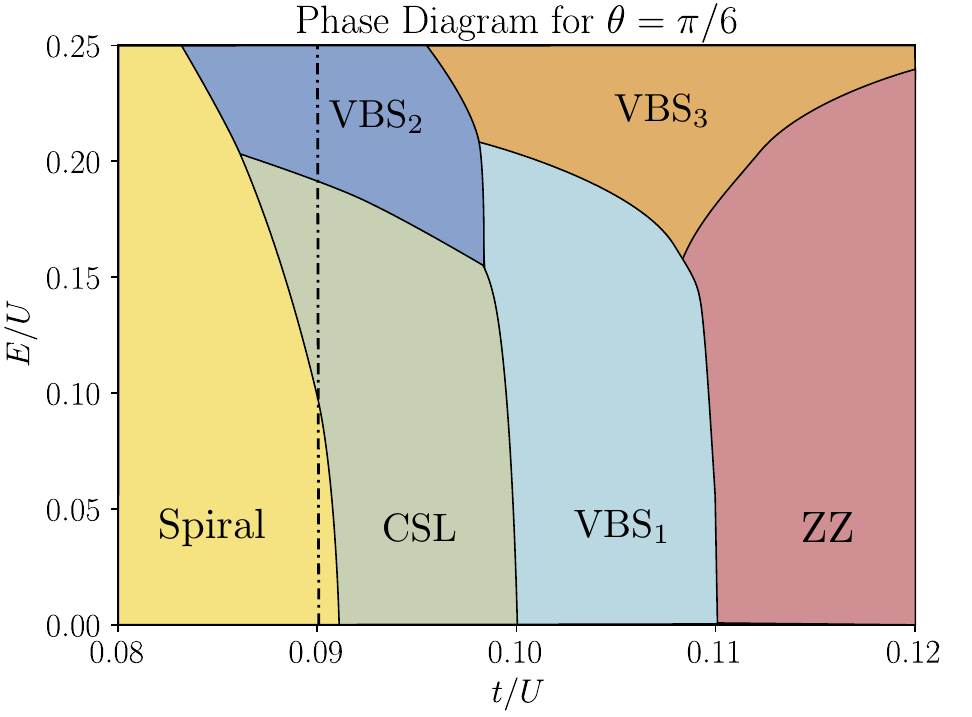}
\caption{Phase diagram with an electric field between the $\mathbf{a}_1$ and $\mathbf{a}_2$ bonds ($\theta = \pi/6$). Observables along the vertical dash-dotted line are depicted in Fig.~\ref{fig:t09_thetapi6_phase}}
\label{fig:phasediagrampi6}
\end{figure}

\begin{figure}
    \centering
    \includegraphics[scale=0.5]{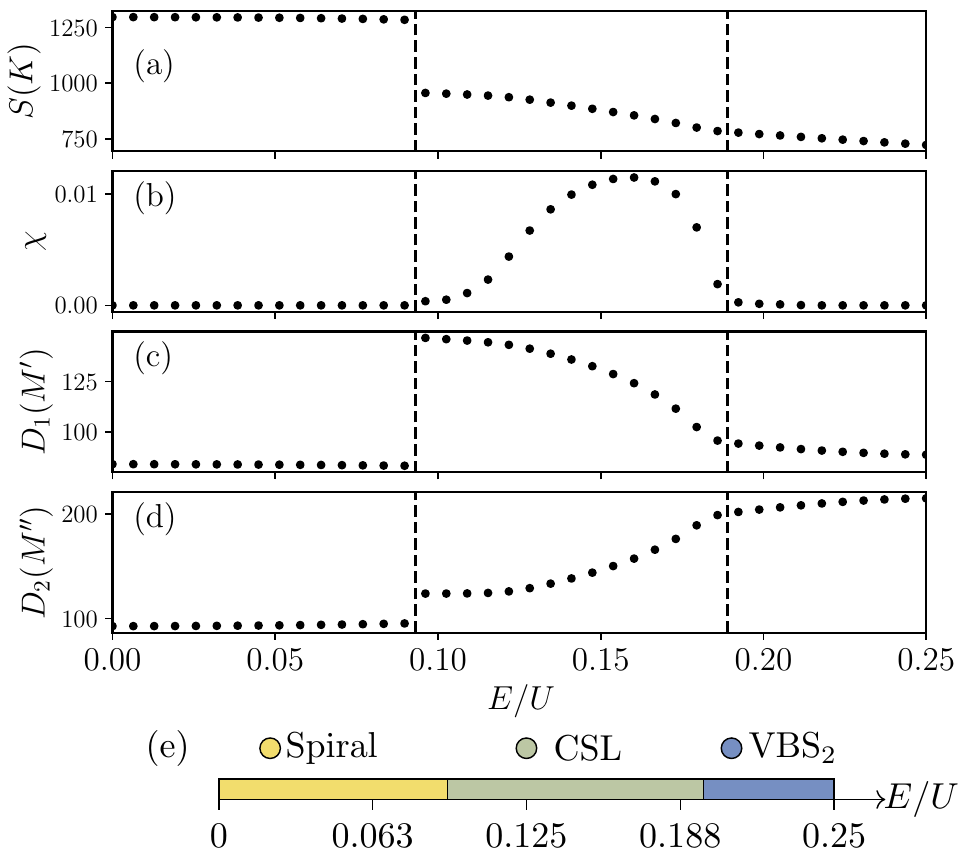}
    \caption{Phase diagram and values of correlators at fixed $t/U = 0.09$ as a function of $E/U$ for the electric field pointing between the $\mathbf{a}_1$ and $\mathbf{a}_2$ bonds ($\theta = \pi/6$). (a) The spin structure factor at the $K$ point in the BZ. (b) The scalar chirality. (c) The dimer structure factor along the $\mathbf{a}_1$ bond at the $M'$ point in the BZ. (d) The dimer structure factor along the $\mathbf{a}_1$ bond at the $M$ point in the BZ. (e) The overall phase diagram for this line}
    \label{fig:t09_thetapi6_phase}
\end{figure}

If we instead align the electric field along $\mathbf{a}_1 + \mathbf{a}_2$ ($\theta = \pi/6$), then the coupling constants evolve in a slightly different way, as can be seen in Fig.~\ref{fig:coupling_e_dependence}. The phase diagram obtained in this configuration is presented in Fig.~\ref{fig:phasediagrampi6}. The same overall behavior as for the $\theta = 0$ case can be observed with the phase boundaries roughly shifting at smaller ratios of $t/U$ as the electric field strength increases. As illustrated by the vertical dash-dotted line in Fig.~\ref{fig:phasediagrampi6} and the evolution of observables along that line in Fig.~\ref{fig:t09_thetapi6_phase}, the electric field can once again be used to stabilize a CSL starting from a spiral ordered phase. For the specific ratio $t/U=0.09$ presented in Fig.~\ref{fig:t09_thetapi6_phase}, one can observe that the electric field necessary to stabilize the CSL is approximately the same as for the $\theta=0$ case (i.e., see Fig.~\ref{fig:t09_thetapi6_phase}). Despite these similarities, some salient differences remain between the phase diagrams obtained for the two field directions. In particular, for large electric fields, the VBS$_2$ phase becomes more stable than the CSL much faster with $\theta=\pi/6$ whereas the CSL ground state persists for the largest values of the electric field examined in the $\theta=0$ case. For $\theta=\pi/6$, there is also a new exotic phase, labeled as VBS$_3$ and discussed below, that emerges for large values of $t/U$ and electric field strength that is absent from the $\theta=0$ phase diagram. These differences between the phase diagrams obtained with the two field directions demonstrate that the electric field direction can be used as a further degree of freedom to potentially promote QSLs and other exotic magnetic phases even for SU(2) symmetric models in the absence of spin-orbit coupling. 

As mentioned above, we find a new phase (VBS$_3$) at a large electric field and ratio of $t/U$, which we tentatively label as a valence bond solid. This state has weak spin-spin correlations that decay fast in real space and rather strong dimer-dimer correlations as seen from the sharp peak in its dimer structure factor illustrated in panels (2.a)-(2.c) of Fig.~\ref{fig:vbs2_vbs3} and the associated slow decay of the dimer-dimer correlations in real space (see Figs.~\ref{fig:sssf_mom_all}-\ref{fig:sdsfc_real_all}). This rules out the possibility of magnetic order and suggests that the phase is once again another VBS state. However, we could not conclusively identify it because the sharp peaks in the dimer structure factor are not at high symmetry points in the Brillouin zone. Thus, we cannot associate the presumed dimer covering with one of the six simple coverings listed in Appendix~\ref{app:dimer_coverings}. At this level of analysis, the most likely scenario is that this phase corresponds to a non-trivial dimer covering with a large unit cell and that the related peaks in its dimer structure factor are at momentum points we cannot access due to our unit cell choice. DMRG simulations could potentially confirm this with a different cylinder circumference $L_y$ that would have access to new points in momentum space. Nevertheless, this phase only appears at large values of $t/U$ and electric field strength that are likely not experimentally relevant. The unambiguous determination of the nature of this phase is also beside the main point of this work (i.e., studying the possibility of stabilizing a QSL by applying an electric field). Further investigation of the nature of this phase is thus left for future work. 

\section{Conclusions}
In summary, we have derived the ring exchange model in the Mott insulating regime of the triangular lattice Hubbard model in the presence of a DC electric field. The resulting model has a nearest-neighbor, next nearest-neighbor, and third nearest-neighbor Heisenberg interactions and ring exchange interactions. For each of the different orientations of bonds and rings, the electric field causes the exchange parameters to be spatially anisotropic. The dependence is quite complicated, but the electric field generally tends to enhance the ring exchange couplings compared to the dominant (nearest-neighbor) Heisenberg interaction. This provides a way of effectively tuning $t/U$ by instead increasing the electric field. 

By performing DMRG simulations, we have computed the phase diagram of the model for two directions of the electric field. Without the electric field, the system magnetically orders at large and small $t/U$ into spiral and zigzag orders, respectively. The system enters a CSL phase or valence bond solid in the intermediate regime. Increasing the electric field along a bond of the lattice, we find that we are able to start in the spiral-ordered phase and enter the CSL. We also find that the VBS switches to a different VBS at a large electric field. If we instead increase the electric field along a direction halfway between two lattice bonds, a magnetically ordered phase can once again be promoted to a CSL. For the two different angle choices, the overall phase diagrams look slightly different and contain different phases, implying that the field direction can be used as an extra degree of freedom to potentially stabilize exotic magnetic phases even in the absence of spin-orbit coupling. 

The physical origin of the CSL shifting towards smaller $t/U$ under increasing $E/U$ may be understood in the following way. Without an electric field, the system's energy is lowered when virtual hopping is possible, which is only allowed by the Pauli principle when the two spins are in opposite directions. The virtual hopping event incurs an energy cost of $U$, and weaker $U$ allows for more frequent charge fluctuations. This explains the origin of the ordinary antiferromagnetic Heisenberg interaction of strength $J = 4t^2/U$. In the presence of an electric field, the energy cost of performing the virtual hopping between sites instead becomes $U \pm \Delta\Phi$, with $\Delta\Phi$ being the change in electric potential energy between the two sites. Hopping with the electric field increases the energy cost of the virtual state, whereas hopping against decreases the energy cost. The average of the two is what gives rise to the nearest-neighbor Heisenberg coupling of
\begin{equation}
J = 2t^2 \left(\frac{1}{U + \Delta\Phi} + \frac{1}{U - \Delta\Phi}\right) = \frac{4t^2}{U}\frac{1}{1 - (\Delta\Phi)^2/U^2},
\end{equation}
which satisfies $J \geq 4t^2/U$ (i.e., this equation is the general form of Eq.~\eqref{eq:j1a} truncated to order $t^2/U$). Thus, enhancement occurs whenever the field has a component along the hopping direction. This reasoning similarly applies to the fourth-order processes we compute in this work. The only additional complication is that the number of virtual hopping paths is vastly increased, but the intuition is still applicable overall. It then makes sense that the ring exchange couplings would increase relative to the nearest-neighbor Heisenberg ones because there may be multiple bonds in the ring that are enhanced by the electric field, whereas the nearest-neighbor Heisenberg has only one such bond. Such a relative enhancement of the ring exchange term is illustrated in Fig.~\ref{fig:coupling_e_dependence}.

It is then of concern to note why the enhanced ring exchange may promote the chiral spin liquid phase. As noted by Cookmeyer et al. ~\cite{cookmeyer_four-spin_2021}, the ring exchange term can be rewritten in terms of chiral terms in the following way. Defining $\mathcal{O}_{\triangleright}(i,j,k) = 2\mathbf{S}_i \cdot (\mathbf{S}_j\times\mathbf{S}_k)$, where $i,j,k$ are oriented counterclockwise around a triangle of orientation $\triangleright$ (defined similarly, also in the counterclockwise direction, for $\mathcal{O}_{\triangleleft}$), and also $\chi_{ijk\ell}^2 = \mathcal{O}_\triangleright(i,j,\ell)\mathcal{O}_{\triangleleft}(k,\ell,j) + \mathcal{O}_{\triangleleft}(k,\ell,j)\mathcal{O}_\triangleright(i,j,\ell)$, then the ring exchange interaction can be rewritten as~\cite{cookmeyer_four-spin_2021}

\begin{align}
R_{ijk\ell} ={}& (\textbf{S}_i\cdot\textbf{S}_j)(\textbf{S}_k\cdot\textbf{S}_\ell) + (\textbf{S}_i\cdot\textbf{S}_\ell)(\textbf{S}_j\cdot\textbf{S}_k) \label{eq:ring_term} \\
& - (\textbf{S}_i\cdot\textbf{S}_k)(\textbf{S}_j\cdot\textbf{S}_\ell) \notag\\
={}& a\chi_{ijk\ell}^2 + b\chi_{ijk\ell}^4 + c\chi_{ijk\ell}^6 + \text{spin bilinears},
\end{align}
for some coefficients $a,b,c$, whose values are not important here.
The connection between ring exchange interactions and chiral symmetry breaking is evident if one performs a mean field decoupling on the above rewriting. This observation, combined with the relative enhancement of the ring exchange, explains why the CSL tends to become favorable as one increases $E/U$. 


We have demonstrated that a CSL can be obtained from a spiral-ordered state by applying a DC electric field to the half-filled triangular lattice Hubbard model at strong coupling. Although one needs to start relatively close to the phase boundary of the CSL, this is an essential proof of concept for the ability of the electric field to act as an experimentally tuning knob to stabilize QSL. It would be interesting to explore the impact of an electric field on the phase diagram of other spin models, where it may have a more pronounced effect. Furthermore, the formalism for deriving the ring exchange model in an electric field applies in principle to any lattice, although the specific forms of couplings would change. Our results present new opportunities for studying field-induced spin liquid behavior. 

\begin{acknowledgements}
We thank Tessa Cookmeyer and Matthias Gohlke for their helpful suggestions. This work was supported by NSERC of Canada and the Center for Quantum Materials at the University of Toronto. Y.B.K. is also supported by the Simons Fellowship from the Simons Foundation and the Guggenheim Fellowship from the John Simon Guggenheim Memorial Foundation. D.S. is supported by an Ontario Graduate Scholarship. F.D. is supported by the Vanier Canada Graduate Scholarship. Computations were performed on the Cedar and Niagara clusters hosted by WestGrid and SciNet in partnership with the Digital Research Alliance of Canada.
\end{acknowledgements}

\appendix

\begin{widetext}

\section{Details of the Canonical Transformation} \label{app:sc_expansion}
The Hubbard Hamiltonian in the electric field is composed of several terms:
\begin{equation}
H = T^+ + T^- + T^0 + H_U + H_E.
\end{equation}
These terms are defined in the main text, but we will repeat their definitions here for convenience:
\begin{align}
T^+ ={}& \sum_{ij} T^+_{ij},\quad T^- = \sum_{ij} T^-_{ij}, \quad T^0 = \sum_{ij} T^0_{ij}, \\
T^+_{ij} ={}& -\sum_\sigma t_{ij} n_{i,-\sigma}c^\dagger_{i\sigma}c_{j\sigma}h_{j,-\sigma}, \\
T^-_{ij} ={}& -\sum_\sigma t_{ij} h_{i,-\sigma} c^\dagger_{i\sigma}c_{j\sigma} n_{j,-\sigma}, \\
T^0_{ij} ={}& -\sum_{\sigma}t_{ij} \left[h_{i,-\sigma}c^\dagger_{i\sigma}c_{j\sigma}h_{j,-\sigma} + n_{i,-\sigma}c^\dagger_{i\sigma}c_{j\sigma}n_{j,-\sigma}\right] \\
H_U ={}& U\sum_i c^\dagger_{i\uparrow} c^\dagger_{i\downarrow} c_{i\downarrow} c_{i\uparrow} \\
H_E ={}& \sum_{i\sigma} \Phi_i c^\dagger_{i\sigma} c_{i\sigma}.
\end{align}
We define the potential to be $V = H_U + H_E$ and the perturbation to be $W = T^+ + T^-$. As mentioned in the main text, we perform a canonical transformation by defining 
\begin{equation}
H_\text{eff} = e^{iS} H e^{-iS}, \label{eq:canonical_transformation_app}
\end{equation}
where we write $S = S^{(1)} + S^{(2)} + S^{(3)} + \cdots$ as an expansion in terms of $S^{(n)} \propto (t/U)^n$. Each term in the expansion is defined so that it will exactly eliminate operators in $H_\text{eff}$, which changes the number of doubly occupied sites to order $t^n/U^{n-1}$. After we solve for $S$ to a certain order $n$, then we project everything to the half-filled subspace. To solve for $S^{(n)}$, we must do an iterative procedure, first finding $S^{(1)}$, then using $S^{(1)}$ to find $S^{(2)}$, and so on. Let us start by expanding Eq.~\eqref{eq:canonical_transformation_app} to order $t^2/U$:
\begin{align}
H_\text{eff}^{(2)} ={}& H + [iS^{(1)}, H] + \frac{1}{2}[iS^{(1)},[iS^{(1)}, H]] \\
={}& T^+ + T^- + T^0 + V + [iS^{(1)},T^+ + T^- + T^0 + V] + \frac{1}{2}[iS^{(1)},[iS^{(1)},T^+ + T^- + T^0 + V]]. \nonumber
\end{align}
To determine $S^{(1)}$, we want to remove operators which change the number of doubly occupied sites. These terms at order $t$ are $T^++T^-$. We see that $[iS^{(1)},V]$ is also of order $t$ and therefore wish to use it to cancel $T^++T^-$, and thereby set 
\begin{equation}
[iS^{(1)}, V] + T^+ + T^- = 0.
\end{equation}
Solving this equation for $iS^{(1)}$ is often done with a guess and check method. However, when such commutator equations become increasingly complicated, it is beneficial to use the following general formula. To solve the commutator equation $[X,A] = B$ for an unknown $X$, we have that 
\begin{equation}
X = i\lim_{\eta \to 0^+}\int_0^\infty e^{iA r} B e^{-iA r} e^{-\eta r} dr.
\end{equation}
Thus, the solution for $iS^{(1)}$ is given by
\begin{equation}
iS^{(1)} = \frac{1}{U}\sum_{ij} \Lambda_{ij}(T^+_{ij} - T^-_{ji}),
\end{equation}
whereby the factor $\Lambda_{ij}$ encodes the electric field:
\begin{equation}
\Lambda_{ij} = \frac{1}{1+\Phi_{ij}}.
\end{equation}
It reduces to one when the electric field is constant, or $\Phi_{ij} = 0$. Note that $S^{(1)}$ reduces to the familiar expression in this case. Now, we can move on to the next order. The effective Hamiltonian to order $t^3/U^2$ is given by using the first two terms in the canonical transformation generator, $iS = iS^{(1)} + iS^{(2)}$:
\begin{align}
H_\text{eff}^{(3)} ={}& H + [iS^{(1)}+iS^{(2)}, H] + \frac{1}{2}[iS^{(1)}+iS^{(2)},[iS^{(1)}+iS^{(2)}, H]] \\
&+ \frac{1}{3!}[iS^{(1)} + iS^{(2)},[iS^{(1)}+iS^{(2)},[iS^{(1)}+iS^{(2)}, H]]]. \nonumber
\end{align}
Truncated to order $t^3/U^2$, this gives us 
\begin{align}
H_\text{eff}^{(3)} ={}& T+V + [iS^{(1)}+iS^{(2)}, T+V] + \frac{1}{2}\left([iS^{(1)},[iS^{(2)},V]] + [iS^{(2)},[iS^{(1)},V]] + [iS^{(1)},[iS^{(1)},T]]\right) \nonumber \\
&+ \frac{1}{3!}[iS^{(1)},[iS^{(1)},[iS^{(1)}, V]]].
\end{align}

We now want to use $iS^{(2)}$ to remove operators which, at order $t^2/U$, bring us away from the half-filled subspace. Expanding out the commutators in the above equation, we have
\begin{align}
H_\text{eff}^{(3)} ={}& T^0 + V + [iS^{(1)},T^0] + \frac{1}{2}[iS^{(1)},T^++T^-] + [iS^{(2)},V] + [iS^{(2)},T] \\
&+\frac{1}{2}\left([iS^{(1)},[iS^{(2)},V]] + [iS^{(2)},[iS^{(1)},V]] + [iS^{(1)},[iS^{(1)},T]]\right) + \frac{1}{3!}[iS^{(1)},[iS^{(1)},[iS^{(1)}, V]]], \nonumber
\end{align}
where the only term that changes the number of doubly occupied sites is $[iS^{(1)},T^0]$, which we will cancel out with the $[iS^{(2)},V]$ term. We can set the sum of these equal to zero and use it to solve for $iS^{(2)}$. Therefore, 
\begin{equation}
[iS^{(1)},T^0] + [iS^{(2)},V] = 0.
\end{equation}
The solution of this commutator equation is given by
\begin{equation}
iS^{(2)} = \frac{1}{U^2}\sum_{ijpq} \Omega_{ijpq}\left([T^+_{ij},T^0_{pq}] + [T^-_{ji},T^0_{qp}]\right).
\end{equation}
The $\Omega_{ijpq}$ factor, containing information about the electric field, is given by
\begin{equation}
\Omega_{ijpq} = \frac{\Lambda_{ij}}{1+\Phi_{ij}/U + \Phi_{pq}/U}.
\end{equation}

Finally, we can calculate the Hamiltonian at fourth order in perturbation theory. We now want to eliminate all operators at $t^3/U^2$ order that leave the singly occupied sector. There are, unfortunately, many more terms that need to be eliminated now. Expanding out the Hamiltonian to 4th order, 
\begin{align}
H_\text{eff}^{(4)} ={}& H + [iS^{(1)}+iS^{(2)} + iS^{(3)}, H] + \frac{1}{2}[iS^{(1)}+iS^{(2)} + iS^{(3)},[iS^{(1)}+iS^{(2)} + iS^{(3)}, H]]  \label{eq:4th_order_expansion} \\
&+ \frac{1}{3!}[iS^{(1)}+iS^{(2)} + iS^{(3)},[iS^{(1)}+iS^{(2)} + iS^{(3)},[iS^{(1)}+iS^{(2)} + iS^{(3)}, H]]] \nonumber \\
&+ \frac{1}{4!}[iS^{(1)}+ iS^{(2)} + iS^{(3)},[iS^{(1)}+iS^{(2)} + iS^{(3)},[iS^{(1)}+iS^{(2)} + iS^{(3)},[iS^{(1)}+iS^{(2)} + iS^{(3)}, H]]]], \nonumber
\end{align}
which, truncated to order $t^4/U^3$ is given by
\begin{align}
H_\text{eff}^{(4)} ={}& H + [iS^{(1)},V] + [iS^{(1)},T] + [iS^{(2)},V] + \frac{1}{2}[iS^{(1)},[iS^{(1)},V]] \\
&+ [iS^{(2)},T] + [iS^{(3)},V] + \frac{1}{2}\left([iS^{(1)},[iS^{(1)},T]] + [iS^{(1)},[iS^{(2)},V]] + [iS^{(2)},[iS^{(1)},V]]\right) \nonumber \\
&+ \frac{1}{3!}[iS^{(1)},[iS^{(1)},[iS^{(1)},V]]] \nonumber \\
&+ [iS^{(3)},T] + \frac{1}{2}\left([iS^{(1)},[iS^{(2)},T]] + [iS^{(2)},[iS^{(1)},T]] + [iS^{(2)},[iS^{(2)},V]] + [iS^{(1)},[iS^{(3)},V]] + [iS^{(3)},[iS^{(1)},V]]\right) \nonumber \\
&+ \frac{1}{3!}\left([iS^{(1)},[iS^{(1)},[iS^{(2)},V]]] + [iS^{(1)},[iS^{(2)},[iS^{(1)},V]]] + [iS^{(2)},[iS^{(1)},[iS^{(1)},V]]] + [iS^{(1)},[iS^{(1)},[iS^{(1)},T]]]\right) \nonumber \\
&+ \frac{1}{4!}[iS^{(1)},[iS^{(1)},[iS^{(1)},[iS^{(1)}, V]]]]. \label{eq:4th_order_truncated}
\end{align}
The first line has order $U,t,t^2/U$ terms, the second and third lines have order $t^3/U^2$ terms, and the last three are order $t^4/U^3$. Of course, there are cancellations of terms that occur due to plugging in the forms of $iS^{(1)},iS^{(2)}$ derived above, but this is simply the general expression. At this stage, we want to remove all operators that change the number of doubly occupied sites to order $t^3/U^2$. The operators in this order which change the number of doubly occupied sites are canceled by the $[iS^{(3)},V]$ term, yielding the condition 
\begin{equation}
[iS^{(3)},V] + [iS^{(2)},T^0] + \frac{1}{2U^2}\sum_{ijpq}\Omega_{ijpq}\left([[T^+_{ij},T^0_{pq}],T^+] + [[T^-_{ji},T^0_{qp}],T^-]\right) + \frac{1}{3}[iS^{(1)},[iS^{(1)},T^++T^-]] = 0
\end{equation}
Solving this equation gives the canonical transformation to be
\begin{align}
iS^{(3)} ={}& \frac{1}{U^3}\sum_{ijpqab} \Xi_{ijpqab} ([[T^+_{ij},T^0_{pq}],T^0_{ab}] - [[T^-_{ji},T^0_{qp}],T^0_{ba}]) + \frac{1}{2U^3}\sum_{ijpqab}\Xi'_{ijpqab}([[T^+_{ij},T^0_{pq}],T^+_{ab}] - [[T^-_{ji},T^0_{qp}],T^-_{ba}]) \\
&+ \frac{1}{3U^3}\sum_{ijpqab}\Lambda_{ij}(\Lambda_{pq} + \Lambda_{ab})\Lambda_{ijpqba}([T^+_{ij},[T^+_{pq},T^-_{ba}]] + [T^-_{ji},[T^+_{ab},T^-_{qp}]])
\end{align}
where the factors encoding the electric field are given by
\begin{align}
\Xi_{ijpqab} ={}& \frac{\Omega_{ijpq}}{1+\Phi_{ij}/U + \Phi_{pq}/U + \Phi_{ab}/U},\\
\Xi'_{ijpqab} ={}& \frac{\Omega_{ijpq}}{2 + \Phi_{ij}/U + \Phi_{pq}/U + \Phi_{ab}/U},\\ \Lambda_{ijpqab} ={}& \frac{1}{1 + \Phi_{ij}/U + \Phi_{pq}/U + \Phi_{ab}/U}.
\end{align}

The effective Hamiltonian, then simplifying Eq.~\eqref{eq:4th_order_truncated} as much as possible without substituting in the specific forms of $iS^{(1)},iS^{(2)}$ and $iS^{(3)}$, gives us 
\begin{align}
H^{(4)}_\text{eff} ={}& T^0 + V + \frac{1}{2U}\sum_{ij}\Lambda_{ij}\left([T^+_{ij},T^-] - [T^-_{ji},T^+]\right) + \frac{1}{2U^2}\sum_{ijpq}\Omega_{ijpq}\left([[T^+_{ij},T^0_{pq}],T^-] + [[T^-_{ji},T^0_{qp}],T^+]\right) \\
&+ [iS^{(3)},T^0] + \frac{1}{2}[iS^{(3)},T^++T^-] + \frac{1}{3}[iS^{(1)},[iS^{(2)},T^++T^-]] +
\frac{1}{3}[iS^{(2)},[iS^{(1)},T^++T^-]] \\
&- \frac{1}{24}[iS^{(1)},[iS^{(1)},[iS^{(1)},T^++T^-]]] - \frac{1}{4U^2}\sum_{ijpq}\Omega_{ijpq}[iS^{(1)},[[T^+_{ij},T^0_{pq}],T^+] + [[T^-_{ji},T^0_{qp}],T^-]].
\end{align}
The resulting Hamiltonian now has exactly zero fluctuations to and from the doubly occupied sector to order $t^3/U^2$. The final step, which is to project to the singly-occupied subspace, thus incurs an error of order $t^4/U^3$. The projection immediately provides numerous simplifications. The key observations are that, since a state $\ket{\Psi}$ in the singly occupied subspace $\mathcal{H}_\text{spin}$ has no doubly occupied sites, so $T^-\ket{\Psi} = 0$ and $\bra{\Psi}T^+ = 0$. Furthermore, $T^0\ket{\Psi} = 0$ because it involves the annihilation of empty and doubly occupied states, neither of which exist in $\mathcal{H}_\text{spin}$. We also note that terms with an unequal number of $T^+$ and $T^-$ operators go to zero because they cannot return to $\mathcal{H}_\text{spin}$. Additionally, $V\ket{\Psi}$ is a constant we can ignore. Due to this, most of the terms in the above disappear completely because they are either constant or every operator in the commutator string has an unequal number of $T^+$ and $T-$. The remaining nontrivial terms then become
\begin{align}
H^{(4)}_\text{eff} ={}&\frac{1}{2U}\sum_{ij}\Lambda_{ij}\left([T^+_{ij},T^-] - [T^-_{ji},T^+]\right) + \frac{1}{2U^2}\sum_{ijpq}\Omega_{ijpq}\left([[T^+_{ij},T^0_{pq}],T^-] + [[T^-_{ji},T^0_{qp}],T^+]\right) \\
&+ \frac{1}{2}[iS^{(3)},T^++T^-] - \frac{1}{24}[iS^{(1)},[iS^{(1)},[iS^{(1)},T^++T^-]]].
\end{align}
Now, plugging the explicit expressions of $iS^{(1)}$ and $iS^{(3)}$ in and relabelling indices appropriately yields
\begin{align}
H^{(4)}_\text{eff} ={}& -\frac{1}{2U}\sum_{ij}(\Lambda_{ij} + \Lambda_{ba})T^-_{ab}T^+_{ij} + \frac{1}{2U^2}\sum_{ijpq}(\Omega_{ijpq} + \Omega_{abqp})T^-_{ab}T^0_{pq}T^+_{ij} \label{eq:effective_ham_2ndQ} \\
&-\frac{1}{2U^3}\sum_{abijmnpq}A_{abijmnpq} T^-_{ab}T^0_{ij}T^0_{mn}T^+_{pq} + \frac{1}{6U^3}\sum_{abijmnpq} B_{abijmnpq} T^-_{ab}T^+_{ij}T^-_{mn}T^+_{pq} \nonumber \\
&- \frac{1}{6U^3}\sum_{abijmnpq} C_{abijmnpq} T^-_{ab}T^-_{ij}T^+_{mn}T^+_{pq} \nonumber \\
&-  \frac{1}{24U^3}\sum_{abijmnpq} D_{abijmnpq} T^-_{ab}T^+_{ij}T^-_{mn}T^+_{pq} + \frac{1}{24U^3}\sum_{abijmnpq} E_{abijmnpq} T^-_{ab}T^-_{ij}T^+_{mn}T^+_{pq} \nonumber
\end{align}
where the electric field-dependent constants are 
\begin{align}
A_{abijmnpq} ={}& \Xi_{bajinm} + \Xi_{pqmnij}, \\
B_{abijmnpq} ={}& (\Lambda_{ba}\Lambda_{ij} + 2\Lambda_{ba}\Lambda_{nm} + \Lambda_{nm}\Lambda_{ij})\Lambda_{nmbaji} + (\Lambda_{pq}\Lambda_{nm} + 2\Lambda_{pq}\Lambda_{ij} + \Lambda_{ij}\Lambda_{nm})\Lambda_{ijpqmn}, \\
C_{abijmnpq} ={}& \Lambda_{ba}(\Lambda_{mn} + \Lambda_{ji})\Lambda_{bajinm} + \Lambda_{pq}(\Lambda_{ji} + \Lambda_{mn})\Lambda_{pqmnij}, \\
D_{abijmnpq} ={}& \Lambda_{ba}\Lambda_{ij}\Lambda_{nm} + 3\Lambda_{ba}\Lambda_{ij}\Lambda_{pq} + \Lambda_{ij}\Lambda_{nm}\Lambda_{pq} + 3\Lambda_{ba}\Lambda_{nm}\Lambda_{pq}, \\
E_{abijmnpq} ={}& 2\Lambda_{ba} \Lambda_{mn} \Lambda_{pq} + 2\Lambda_{ba} \Lambda_{ji} \Lambda_{pq}.
\end{align}

The last step is to convert the strings of second quantized operators into the spin Hilbert space. To do this, one can compute the matrix elements of the $H_\text{eff}^{(4)}$ in the basis of spin up/down on every site. This will yield bilinear terms across nearest-neighbor, next nearest-neighbor, and third nearest-neighbor sites, as well as ring exchange terms that connect four spins in a ring. Three spin terms are prohibited by time-reversal symmetry, which the electric field does not break. This gives the result Eq.~\eqref{eq:spin_model} in the main text, with coupling constants given in Appendix~\ref{app:couplings}. Explicitly, if we have some operator $\mathcal{O}$ which acts on $N$ sites (in the singly-occupied subspace), then we wish to express it in terms of spin operators as
\begin{equation}
\mathcal{O} = \sum_{i_1,\dots,i_N=x,y,z} c_{i_1,\dots,i_N} S^{i_1}_1\dots S^{i_N}_N,
\end{equation}
where $S^i_n$ is the $i$th component ($i=x,y,z$) of a spin operator acting on site $n$. To do this, we note that the strings of fermionic operators appear in two forms, both of which are related to spin operators:
\begin{align}
h_{-\alpha} c^\dagger_\alpha c_\beta h_{-\beta} ={}& \ket{\alpha}\bra{\beta} = \delta_{\alpha\beta}(S^0 + \eta(\alpha)S^z) + \delta_{\alpha,-\beta}(S^x + i\eta(\alpha)S^y), \\
c_{\alpha}n_{-\alpha}n_{-\beta}c^\dagger_\beta ={}& \eta(\alpha)\eta(\beta)\ket{-\alpha}\bra{-\beta} = \eta(\beta)[\delta_{\alpha\beta}(S^0\eta(\alpha) - S^z) + \delta_{\alpha,-\beta}(S^x\eta(\alpha) - iS^y)],
\end{align}
where $\eta(\uparrow) = 1$, $\eta(\downarrow) = -1$, and the $S$ operators are half of the Pauli matrices (including $S^0$, which is half of the $2\times 2$ identity matrix). Note that when putting operators into the above form, one needs to be careful of fermionic exchange statistics. Doing this leads to the spin model in Eq.~\eqref{eq:spin_model}.

\section{Electric-field dependent coupling constants} \label{app:couplings}
We take the matrix elements of the effective Hamiltonian in Eq.~\eqref{eq:effective_ham_2ndQ} on different numbers of sites, which essentially leads to different site indices being set equal to one another because the electrons must undergo virtual hoppings in a closed loop. For the second order ($t^2/U$) contribution, the effective Hamiltonian becomes 
\begin{align}
H^{(2)} = \frac{2}{U}\sum_{ij}t_{ij}^2\ \frac{1}{1-\Phi_{ij}^2/U^2} \textbf{S}_i\cdot \textbf{S}_j
\end{align}
The order $t^3/U$ contribution yields nothing because time-reversal symmetry is not broken. The terms at order $t^4/U^3$ have numerous contributions. First, there is the case when the electrons undergo 4 hopping processes but just on two sites, which leads to only nearest-neighbor interactions. Then, there are contributions where the electrons hop 4 times, but this time on 3 different sites. If these three sites are collinear, this leads to third nearest-neighbor interactions, but if they are not, it leads to second nearest-neighbor interactions. Both of these also contain nearest-neighbor terms as well. Lastly, the contributions where electrons hop on 4 sites in a closed loop gives contributions on a rhombus, which yields terms for nearest-neighbor, 2nd nearest-neighbor, and ring exchange couplings. It is somewhat involved to obtain these electric field dependent coupling constants. Thus, we will show how it is done in detail for the two site contributions, and then show only a couple intermediate steps for the three and four site contributions.

\subsection{Two site contributions}

At fourth order, there are three operator strings to consider: $T^-T^0T^0T^+$, $T^-T^+T^-T^+$, and $T^-T^-T^+T^+$. For the two site contribution, the operator string $T^-T^0T^0T^+$ first creates a doubly occupied site, meaning one of the two sites is empty and the other is full. From here, there is no way to move a single electron without changing the number of doubly occupied sites (which is the role of $T^0$), which tells us that $(T^-T^0T^0T^+)_\text{2-site} = 0$. Furthermore, it is not possible to create two doubly occupied sites if you only have two sites to work with, so also $(T^-T^-T^+T^+)_\text{2-site} = 0$. Thus, we only need to find the representation of $T^-T^+T^-T^+$ on two sites. We therefore overlap the 8 sites to get two nonzero terms, when projected to the singly occupied subspace:
\begin{align}
\left(T^-_{ab} T^+_{ij} T^-_{mn} T^+_{pq}\right)_\text{2-site} =& - 4\delta_{bi}\delta_{aj}\delta_{np}\delta_{mq} \delta_{am} \delta_{nb} t_{ba}^4 \textbf{S}_a\cdot\textbf{S}_b - 4\delta_{bi}\delta_{aj}\delta_{np}\delta_{mq} \delta_{an} \delta_{bm} t_{ba}^4 \textbf{S}_a\cdot\textbf{S}_b. \label{eq:2site_projection}
\end{align}
Note that, strictly speaking, taking matrix elements of an operator such as $T^-_{ji}T^+_{ij}T^-_{ji}T^+_{ij}$, we would half
\begin{align}
T^-_{ji}T^+_{ij}T^-_{ji}T^+_{ij} ={}& 4t_{ij}^4(S^0_i S^0_j - \mathbf{S}_i \cdot \mathbf{S}_j)
\end{align}
where $S^0 = \sigma^0/2$ is half of the identity matrix, which only leads to a constant offset in energy and is therefore discarded. Plugging Eq.~\eqref{eq:2site_projection} into Eq.~\eqref{eq:effective_ham_2ndQ} therefore gives us 
\begin{align}
H^{(4)}_\text{2-site} ={}& \sum_{ij}t_{ij}^4\left[\frac{1}{6U^3}(B_{jiijjiij} +  B_{jiijijji})(-4\mathbf{S}_i\cdot\mathbf{S}_j) - \frac{1}{24U^3}(D_{jiijjiij} +  D_{jiijijji})(-4\mathbf{S}_i\cdot\mathbf{S}_j) \right] \\
={}& -\frac{2}{U^3}\sum_{ij} t_{ij}^4 \Lambda_{ij}(2\Lambda_{ij}^2 + \Lambda_{ij}\Lambda_{ji} + \Lambda_{ji}^2)\mathbf{S}_i \cdot\mathbf{S}_j
\end{align}
Since the hopping is symmetric, we can symmetrize the above expression to get 
\begin{align}
H^{(4)}_\text{2-site} ={}& -\frac{1}{U^3}\sum_{ij} t_{ij}^4 [\Lambda_{ij}(2\Lambda_{ij}^2 + \Lambda_{ij}\Lambda_{ji} + \Lambda_{ji}^2) + \Lambda_{ji}(2\Lambda_{ji}^2 + \Lambda_{ji}\Lambda_{ij} + \Lambda_{ij}^2)]\mathbf{S}_i \cdot\mathbf{S}_j \\
={}& -\frac{8}{U^3} \sum_{ij} t_{ij}^4  \frac{1+\Phi_{ij}^2/U^2}{(1-\Phi_{ij}^2/U^2)^3} \textbf{S}_i\cdot\textbf{S}_j.
\end{align}

\subsection{Three site contributions}
We again consider the three operator strings: $T^-T^0T^0T^+$, $T^-T^+T^-T^+$, and $T^-T^-T^+T^+$. On three sites, again it is not possible to create two doubly occupied sites (we only have 3 electrons total, because the system is half-filled), so we have that $(T^-T^-T^+T^+)_\text{3-site} = 0$. We thus need to consider the other two terms. These two terms yield 

\begin{align}
\left(-T^-_{ab} T^0_{ij} T^0_{mn} T^+_{pq}\right)_\text{3-site} ={}& 2\delta_{aj} \delta_{in} \delta_{mq} \delta_{bp} \delta_{am} t_{pa}^2 t_{ai}^2\left(\textbf{S}_p \cdot\textbf{S}_a\right) + 2\delta_{bi}\delta_{jm}\delta_{aq}\delta_{np} \delta_{pi} t_{ai}^2 t_{im}^2 \left(
\textbf{S}_a \cdot \textbf{S}_i\right) \\
& +  \delta_{bi}\delta_{an}\delta_{jp}\delta_{mq} \delta_{im} t_{am}^2 t_{mp}^2 \left(\textbf{S}_m \cdot \textbf{S}_p - \textbf{S}_a \cdot\textbf{S}_p + \textbf{S}_a \cdot\textbf{S} _m\right)  \notag \\
& + \delta_{aj}\delta_{bm}\delta_{iq}\delta_{pn} \delta_{ap} t_{ma}^2 t_{ai}^2 \left(\textbf{S}_a \cdot \textbf{S}_i  - \textbf{S}_m \cdot\textbf{S}_i + \textbf{S}_m \cdot \textbf{S}_a\right), \notag
\end{align}
\begin{align}
\left(T^-_{ab} T^+_{ij} T^-_{mn} T^+_{pq}\right)_\text{3-site} ={}&  \delta_{bi}\delta_{aj}\delta_{np}\delta_{mq} \delta_{am} t_{ba}^2 t_{an}^2 \left(- \textbf{S}_a \cdot \textbf{S}_n + \textbf{S}_b \cdot \textbf{S}_n -\textbf{S}_b \cdot \textbf{S}_a\right)  \notag \\
&+ \delta_{bi}\delta_{aj}\delta_{np}\delta_{mq} \delta_{an} t_{ba}^2 t_{am} \left( - \textbf{S}_a \cdot \textbf{S}_m + \textbf{S}_b \cdot \textbf{S}_m - \textbf{S}_b \cdot \textbf{S}_a\right)  \notag \\
&+ \delta_{bi}\delta_{aj}\delta_{np}\delta_{mq} \delta_{bm} t_{ab}^2 t_{bn}^2 \left(
-\textbf{S}_b \cdot \textbf{S}_n + \textbf{S}_a \cdot \textbf{S}_n -  \textbf{S}_a \cdot \textbf{S}_b\right)  \notag \\
&+ \delta_{bi}\delta_{aj}\delta_{np}\delta_{mq} \delta_{bn} t_{ab}^2 t_{bm}^2 \left( -\textbf{S}_b \cdot \textbf{S}_m + \textbf{S}_a \cdot \textbf{S}_m - \textbf{S}_a \cdot \textbf{S}_b\right),  \notag
\end{align}
where the delta functions cause the sites to overlap. The delta functions are explicitly kept to illustrate how the site indices on the $A,B,C,D,E$ coefficients need to be replaced as the different overlapping configurations are considered. As in the two site case, the resulting $A,B,C,D,E$ coefficients will simplify when the symmetrized combinations are taken. Doing this gives the three site Hamiltonian to be
\begin{align}
H^{(4)}_\text{3-site} ={}& \frac{2}{U^3} \sum_{ijk}  t_{ij}^2 t_{jk}^2\left\{\textbf{S}_i \cdot\textbf{S}_k \left[2\frac{1-\Phi_{ij}^2\Phi_{jk}^2/U^4}{(1-\Phi_{i j}^2/U^2)^2(1-\Phi_{j k}^2/U^2)^2} -\frac{1 + \Phi_{ij}\Phi_{ik}/U^2 + \Phi_{ij}\Phi_{jk}/U^2 + \Phi_{ik}\Phi_{jk}/U^2}{(1-\Phi_{ij}^2/U^2)(1-\Phi_{jk}^2/U^2)(1-\Phi_{ik}^2/U^2)}\right] \right. \\
&+ 2\textbf{S}_i \cdot\textbf{S}_j\left[\frac{1+2\Phi_{ij}\Phi_{ik}/U^2 + \Phi_{ij}^2/U^2}{(1-\Phi_{ij}^2/U^2)^2(1-\Phi_{ik}^2/U^2)} - \frac{2(1-\Phi_{ij}^2\Phi_{jk}^2/U^4)}{(1-\Phi_{i j}^2/U^2)^2(1-\Phi_{j k}^2/U^2)^2} \right.  \notag \\
&\left.\left.+ \frac{1 + \Phi_{ij}\Phi_{ik}/U^2 + \Phi_{ij}\Phi_{jk}/U^2 + \Phi_{ik}\Phi_{jk}/U^2}{(1-\Phi_{ij}^2/U^2)(1-\Phi_{jk}^2/U^2)(1-\Phi_{ik}^2/U^2)}\right] \right\}. \notag
\end{align}
The reason that this three-site contribution yields nearest-neighbor interactions is clear from the $\mathbf{S}_i\cdot\mathbf{S}_j$ term, but in fact, the $\mathbf{S}_i\cdot\mathbf{S}_k$ term can also contribute to the nearest-neighbor interaction on the triangular lattice (but not on the square lattice). Finally, the $\mathbf{S}_i\cdot\mathbf{S}_k$ term contributes to both the second and third nearest-neighbor interactions because three sites participate in virtual hoppings connecting those spins. 

\subsection{Four site contributions}
The three operator strings $T^-T^0T^0T^+$, $T^-T^+T^-T^+$, and $T^-T^-T^+T^+$ all contribute to the four site Hamiltonian. This time however, the $T^-T^+T^-T^+$ term only produces ``disconnected" terms, which have intermediate hoppings of the form $t_{ij}^2 t_{k\ell}^2$. These disconnected terms are cancelled by further terms in $T^-T^-T^+T^+$. In the below, we will only list the connected terms:
\begin{align}
\left(-T^-_{ab} T^0_{ij} T^0_{mn} T^+_{pq}\right)_\text{4-site}  ={}& \frac{1}{2}t_{ap}t_{pm}t_{mi}t_{ia} \delta_{aj} \delta_{in} \delta_{mq} \delta_{bp} \left(4R_{apmi} -  \textbf{S}_m \cdot\textbf{S}_i + \textbf{S}_p \cdot \textbf{S}_i  \right. \\
&\left. + \textbf{S}_p\cdot\textbf{S}_m - \textbf{S}_a \cdot \textbf{S}_i - \textbf{S}_a \cdot  \textbf{S}_m + \textbf{S}_a \cdot  \textbf{S}_p \right) \notag \\
&+\frac{1}{2} t_{ai}t_{ip}t_{pm}t_{ma}\delta_{bi}\delta_{an}\delta_{jp}\delta_{mq} \left(4R_{aipm} - \textbf{S}_p \cdot\textbf{S}_m - \textbf{S}_i \cdot \textbf{S}_m \right. \notag \\
&\left. + \textbf{S}_i \cdot\textbf{S}_p + \textbf{S}_a \textbf{S}_m - \textbf{S}_a \cdot \textbf{S}_p  - \textbf{S}_a \cdot\textbf{S}_i \right) \notag \\
&+ \frac{1}{2}t_{am}t_{mp}t_{pi}t_{pa} \delta_{aj}\delta_{bm}\delta_{iq}\delta_{pn} \left( 4R_{ampi} - \textbf{S}_p \cdot\textbf{S}_i - \textbf{S}_m \cdot \textbf{S}_i \right. \notag \\
&\left. + \textbf{S}_m \cdot\textbf{S}_p + \textbf{S}_a \cdot \textbf{S}_i - \textbf{S}_a \textbf{S}_p  - \textbf{S}_a \cdot\textbf{S}_m \right) \notag \\
&+ \frac{1}{2} t_{ai} t_{im} t_{mp} t_{pa} \delta_{bi}\delta_{jm}\delta_{aq}\delta_{np} \left(4R_{aimp} - \textbf{S}_m \cdot\textbf{S}_p - \textbf{S}_i \cdot \textbf{S}_p  \right. \notag \\
&\left. -\textbf{S}_i \cdot\textbf{S}_m + \textbf{S}_a \textbf{S}_p + \textbf{S}_a \textbf{S}_m + \textbf{S}_a \cdot\textbf{S}_i \right), \notag 
\end{align}
and the ring exchange contribution (ignoring the disconnected part) from the $T^-T^-T^+T^+$ term is given by 
\begin{align}
\left(-T^-_{ab} T^-_{ij} T^+_{mn} T^+_{pq}\right)_\text{4-site} ={}& \frac{1}{2} t_{am} t_{mi} t_{ip}t_{pa} \delta_{qa}\delta_{pj}\delta_{in}\delta_{bm} \left(4R_{amip} - \textbf{S}_i \cdot\textbf{S}_p +  \textbf{S}_m \cdot \textbf{S}_p \right. \\
&\left.  - \textbf{S}_m \cdot\textbf{S}_i - \textbf{S}_a \cdot \textbf{S}_p + \textbf{S}_a \cdot \textbf{S}_i - \textbf{S}_a \cdot\textbf{S}_m\right) \notag \\
&+ \frac{1}{2} t_{ap}t_{pi}t_{im}t_{ma} \delta_{qi}\delta_{pb}\delta_{an}\delta_{jm} \left(4R_{apim} -  \textbf{S}_i \cdot\textbf{S}_m +  \textbf{S}_p \cdot \textbf{S}_m\right. \notag \\
&\left.  - \textbf{S}_p \cdot\textbf{S}_i - \textbf{S}_a \cdot \textbf{S}_m + \textbf{S}_a \cdot \textbf{S}_i - \textbf{S}_a \cdot\textbf{S}_p \right). \notag
\end{align}
Here, $R_{amip}$ is as defined in Eq.~\eqref{eq:ring_term}. Plugging these in and applying the $\delta$-functions, and also symmetrizing everything, we find that we have, first, a ring exchange term:
\begin{align}
H^\text{ring}_\text{4-site} ={}& \frac{1}{4U^3} \sum_{ijk\ell} t_{ij} t_{jk} t_{k\ell}t_{\ell i} \left((\textbf{S}_i\cdot\textbf{S}_j)(\textbf{S}_k\cdot\textbf{S}_\ell) + (\textbf{S}_i\cdot\textbf{S}_\ell)(\textbf{S}_j\cdot\textbf{S}_k) - (\textbf{S}_i\cdot\textbf{S}_k)(\textbf{S}_j\cdot\textbf{S}_\ell)\right) \\
& \times \left\{ 4\frac{1-\Phi_{ij}\Phi_{j\ell}/U^2-\Phi_{jk}\Phi_{ij}/U^2+\Phi_{j\ell}\Phi_{jk}/U^2}{(1-\Phi_{ij}^2/U^2)(1-\Phi_{j\ell}^2/U^2)(1-\Phi_{jk}^2/U^2)} + 4\frac{1+\Phi_{ij}\Phi_{ik}/U^2 - \Phi_{ij}\Phi_{k\ell}/U^2 - \Phi_{ik}\Phi_{k\ell}/U^2}{(1-\Phi_{ij}^2/U^2)(1-\Phi_{ik}^2/U^2)(1-\Phi_{k\ell}^2/U^2)} \right.  \notag \\
&+ 4\frac{1-\Phi_{ij}\Phi_{j\ell}/U^2 - \Phi_{ij}\Phi_{k\ell}/U^2 + \Phi_{j\ell}\Phi_{k\ell}/U^2}{(1-\Phi_{ij}^2/U^2)(1-\Phi_{j\ell}^2/U^2)(1-\Phi_{k\ell}^2/U^2)} + 4\frac{1+\Phi_{ij}\Phi_{ik}/U^2 + \Phi_{ij}\Phi_{i\ell}/U^2 + \Phi_{ik}\Phi_{i\ell}/U^2}{(1-\Phi_{ij}^2/U^2)(1-\Phi_{ik}^2/U^2)(1-\Phi_{i\ell}^2/U^2)}  \notag \\
&+ 4\frac{1-\Phi_{jk}\Phi_{ki}/U^2-\Phi_{k\ell }\Phi_{jk}/U^2+\Phi_{ki}\Phi_{k\ell }/U^2}{(1-\Phi_{jk}^2/U^2)(1-\Phi_{ki}^2/U^2)(1-\Phi_{k\ell }^2/U^2)} + 4\frac{1+\Phi_{jk}\Phi_{j\ell }/U^2 - \Phi_{jk}\Phi_{\ell i}/U^2 - \Phi_{j\ell }\Phi_{\ell i}/U^2}{(1-\Phi_{jk}^2/U^2)(1-\Phi_{j\ell }^2/U^2)(1-\Phi_{\ell i}^2/U^2)}  \notag \\
&+ 4\frac{1-\Phi_{jk}\Phi_{ki}/U^2 - \Phi_{jk}\Phi_{\ell i}/U^2 + \Phi_{ki}\Phi_{\ell i}/U^2}{(1-\Phi_{jk}^2/U^2)(1-\Phi_{ki}^2/U^2)(1-\Phi_{\ell i}^2/U^2)} + 4\frac{1-\Phi_{k\ell }\Phi_{\ell j}/U^2 -\Phi_{\ell i }\Phi_{k\ell }/U^2 + \Phi_{\ell j}\Phi_{\ell i }/U^2}{(1-\Phi_{k\ell }^2/U^2)(1-\Phi_{\ell j}^2/U^2)(1-\Phi_{\ell i }^2/U^2)}  \notag \\
&+3\frac{1+\Phi_{ij}\Phi_{k\ell}/U^2 - \Phi_{ij}\Phi_{\ell i}/U^2 - \Phi_{k\ell}\Phi_{\ell i}/U^2}{(1-\Phi_{ij}^2/U^2)(1-\Phi_{k\ell}^2/U^2)(1-\Phi_{\ell i}^2/U^2)} - \frac{1 +\Phi_{ij}\Phi_{k\ell}/U^2 - \Phi_{ij}\Phi_{jk}/U^2 - \Phi_{k\ell}\Phi_{jk}/U^2}{(1-\Phi_{ij}^2/U^2)(1-\Phi_{k\ell}^2/U^2)(1-\Phi_{jk}^2/U^2)}  \notag \\
&+3\frac{1-\Phi_{ij}\Phi_{jk}/U^2 - \Phi_{ij}\Phi_{\ell i}/U^2 + \Phi_{jk}\Phi_{\ell i}/U^2}{(1-\Phi_{ij}^2/U^2)(1-\Phi_{jk}^2/U^2)(1-\Phi_{\ell i}^2/U^2)} - \frac{1 - \Phi_{k\ell}\Phi_{jk}/U^2 - \Phi_{k\ell}\Phi_{\ell i}/U^2 + \Phi_{jk}\Phi_{\ell i}/U^2}{(1-\Phi_{k\ell}^2/U^2)(1-\Phi_{jk}^2/U^2)(1-\Phi_{\ell i}^2/U^2)}  \notag \\
&+ 3\frac{1 - \Phi_{jk}\Phi_{k\ell}/U^2 - \Phi_{jk}\Phi_{ij}/U^2 + \Phi_{k\ell}\Phi_{ij}/U^2}{(1-\Phi_{jk}^2/U^2)(1-\Phi_{k\ell}^2/U^2)(1-\Phi_{ij}^2/U^2)} - \frac{1 - \Phi_{\ell i}\Phi_{k\ell}/U^2 - \Phi_{\ell i}\Phi_{ij}/U^2 + \Phi_{k\ell}\Phi_{ij}/U^2}{(1-\Phi_{\ell i}^2/U^2)(1-\Phi_{k\ell}^2/U^2)(1-\Phi_{ij}^2/U^2)}  \notag \\
&\left. + 3\frac{1-\Phi_{k\ell}\Phi_{\ell i}/U^2 - \Phi_{k\ell}\Phi_{jk}/U^2 + \Phi_{\ell i}\Phi_{jk}/U^2}{(1-\Phi_{k\ell}^2/U^2)(1-\Phi_{\ell i}^2/U^2)(1-\Phi_{jk}^2/U^2)} - \frac{1-\Phi_{ij}\Phi_{\ell i}/U^2 - \Phi_{ij}\Phi_{jk}/U^2 + \Phi_{\ell i}\Phi_{jk}/U^2}{(1-\Phi_{ij}^2/U^2)(1-\Phi_{\ell i}^2/U^2)(1-\Phi_{jk}^2/U^2)} \right\}. \notag
\end{align}
There are two types of spin bilinear Hamiltonians. The first contributes only to the nearest neighbor interaction:
\begin{align*}
H^\text{bilin,1}_\text{4-site} ={}& \frac{1}{4U^3} \sum_{ijk\ell} t_{ij} t_{jk} t_{k\ell}t_{\ell i} \textbf{S}_i\cdot\textbf{S}_j \\
&\times \left\{4\frac{1-\Phi_{ij}\Phi_{j\ell}/U^2 - \Phi_{jk}\Phi_{ij}/U^2 + \Phi_{j\ell}\Phi_{jk}/U^2}{(1-\Phi_{ij}^2/U^2)(1-\Phi_{j\ell}^2/U^2)(1-\Phi_{jk}^2/U^2)}  -4\frac{1+\Phi_{ij}\Phi_{ik}/U^2 - \Phi_{ij}\Phi_{k\ell}/U^2 - \Phi_{ik}\Phi_{k\ell}/U^2}{(1-\Phi_{ij}^2/U^2)(1-\Phi_{ik}^2/U^2)(1-\Phi_{k\ell}^2/U^2)} \right.\\
& -4\frac{1-\Phi_{ij}\Phi_{j\ell}/U^2 - \Phi_{ij}\Phi_{k\ell}/U^2 + \Phi_{j\ell}\Phi_{k\ell}/U^2}{(1-\Phi_{ij}^2/U^2)(1-\Phi_{j\ell}^2/U^2)(1-\Phi_{k\ell}^2/U^2)} -4\frac{1-\Phi_{jk}\Phi_{ki}/U^2 - \Phi_{k\ell }\Phi_{jk}/U^2 + \Phi_{ki}\Phi_{k\ell }/U^2 }{(1-\Phi_{jk}^2/U^2)(1-\Phi_{ki}^2/U^2)(1-\Phi_{k\ell }^2/U^2)} \\
&+ 4\frac{1+\Phi_{jk}\Phi_{j\ell }/U^2 - \Phi_{jk}\Phi_{\ell i}/U^2 - \Phi_{j\ell }\Phi_{\ell i}/U^2}{(1-\Phi_{jk}^2/U^2)(1-\Phi_{j\ell }^2/U^2)(1-\Phi_{\ell i}^2/U^2)} + 4\frac{1-\Phi_{jk}\Phi_{ki}/U^2 - \Phi_{jk}\Phi_{\ell i}/U^2 + \Phi_{ki}\Phi_{\ell i}/U^2}{(1-\Phi_{jk}^2/U^2)(1-\Phi_{ki}^2/U^2)(1-\Phi_{\ell i}^2/U^2)} \\
& -4\frac{1-\Phi_{k\ell }\Phi_{\ell j}/U^2 - \Phi_{\ell i }\Phi_{k\ell }/U^2 + \Phi_{\ell j}\Phi_{\ell i }/U^2}{(1-\Phi_{k\ell }^2/U^2)(1-\Phi_{\ell j}^2/U^2)(1-\Phi_{\ell i }^2/U^2)} +4\frac{1-\Phi_{\ell i }\Phi_{i k}/U^2 - \Phi_{i j }\Phi_{\ell i }/U^2 + \Phi_{i k}\Phi_{i j }/U^2}{(1-\Phi_{\ell i }^2/U^2)(1-\Phi_{i k}^2/U^2)(1-\Phi_{i j }^2/U^2)} \\
& -3\frac{1+\Phi_{ij}\Phi_{k\ell}/U^2 - \Phi_{ij}\Phi_{\ell i}/U^2 - \Phi_{k\ell}\Phi_{\ell i}/U^2}{(1-\Phi_{ij}^2/U^2)(1-\Phi_{k\ell}^2/U^2)(1-\Phi_{\ell i}^2/U^2)} + \frac{1 +\Phi_{ij}\Phi_{k\ell}/U^2 - \Phi_{ij}\Phi_{jk}/U^2 - \Phi_{k\ell}\Phi_{jk}/U^2}{(1-\Phi_{ij}^2/U^2)(1-\Phi_{k\ell}^2/U^2)(1-\Phi_{jk}^2/U^2)} \\
&-3\frac{1-\Phi_{ij}\Phi_{jk}/U^2 - \Phi_{ij}\Phi_{\ell i}/U^2 + \Phi_{jk}\Phi_{\ell i}/U^2}{(1-\Phi_{ij}^2/U^2)(1-\Phi_{jk}^2/U^2)(1-\Phi_{\ell i}^2/U^2)} + \frac{1 - \Phi_{k\ell}\Phi_{jk}/U^2 - \Phi_{k\ell}\Phi_{\ell i}/U^2 + \Phi_{jk}\Phi_{\ell i}/U^2}{(1-\Phi_{k\ell}^2/U^2)(1-\Phi_{jk}^2/U^2)(1-\Phi_{\ell i}^2/U^2)} \\
&- 3\frac{1 - \Phi_{jk}\Phi_{k\ell}/U^2 - \Phi_{jk}\Phi_{ij}/U^2 + \Phi_{k\ell}\Phi_{ij}/U^2}{(1-\Phi_{jk}^2/U^2)(1-\Phi_{k\ell}^2/U^2)(1-\Phi_{ij}^2/U^2)} + \frac{1 - \Phi_{\ell i}\Phi_{k\ell}/U^2 - \Phi_{\ell i}\Phi_{ij}/U^2 + \Phi_{k\ell}\Phi_{ij}/U^2}{(1-\Phi_{\ell i}^2/U^2)(1-\Phi_{k\ell}^2/U^2)(1-\Phi_{ij}^2/U^2)} \\
&\left. - 3\frac{1-\Phi_{k\ell}\Phi_{\ell i}/U^2 - \Phi_{k\ell}\Phi_{jk}/U^2 + \Phi_{\ell i}\Phi_{jk}/U^2}{(1-\Phi_{k\ell}^2/U^2)(1-\Phi_{\ell i}^2/U^2)(1-\Phi_{jk}^2/U^2)} + \frac{1-\Phi_{ij}\Phi_{\ell i}/U^2 - \Phi_{ij}\Phi_{jk}/U^2 + \Phi_{\ell i}\Phi_{jk}/U^2}{(1-\Phi_{ij}^2/U^2)(1-\Phi_{\ell i}^2/U^2)(1-\Phi_{jk}^2/U^2)} \right\}.
\end{align*}
There is another spin bilinear term which connects either nearest-neighbor or next nearest-neighbor spins, depending on how $i,j,k,\ell$ are chosen around a ring:
\begin{align}
H^\text{bilin,2}_\text{4-site} ={}& \frac{1}{8U^3} \sum_{ijk\ell} t_{ij} t_{jk} t_{k\ell}t_{\ell i}  \textbf{S}_i\cdot\textbf{S}_k \\
&\times\left\{-4\frac{1-\Phi_{ij}\Phi_{j\ell}/U^2 - \Phi_{jk}\Phi_{ij}/U^2 + \Phi_{j\ell}\Phi_{jk}/U^2}{(1-\Phi_{ij}^2/U^2)(1-\Phi_{j\ell}^2/U^2)(1-\Phi_{jk}^2/U^2)} -4\frac{1+\Phi_{ij}\Phi_{ik}/U^2 - \Phi_{ij}\Phi_{k\ell}/U^2 - \Phi_{ik}\Phi_{k\ell}/U^2}{(1-\Phi_{ij}^2/U^2)(1-\Phi_{ik}^2/U^2)(1-\Phi_{k\ell}^2/U^2)} \right.  \notag \\
& -4\frac{1-\Phi_{ij}\Phi_{j\ell}/U^2 - \Phi_{ij}\Phi_{k\ell}/U^2 + \Phi_{j\ell}\Phi_{k\ell}/U^2}{(1-\Phi_{ij}^2/U^2)(1-\Phi_{j\ell}^2/U^2)(1-\Phi_{k\ell}^2/U^2)} +4\frac{1-\Phi_{jk}\Phi_{ki}/U^2 - \Phi_{k\ell }\Phi_{jk}/U^2 + \Phi_{ki}\Phi_{k\ell }/U^2}{(1-\Phi_{jk}^2/U^2)(1-\Phi_{ki}^2/U^2)(1-\Phi_{k\ell }^2/U^2)}  \notag \\
& -4\frac{1+\Phi_{jk}\Phi_{j\ell }/U^2 - \Phi_{jk}\Phi_{\ell i}/U^2 - \Phi_{j\ell }\Phi_{\ell i}/U^2}{(1-\Phi_{jk}^2/U^2)(1-\Phi_{j\ell }^2/U^2)(1-\Phi_{\ell i}^2/U^2)}  -4\frac{1-\Phi_{jk}\Phi_{ki}/U^2 - \Phi_{jk}\Phi_{\ell i}/U^2 + \Phi_{ki}\Phi_{\ell i}/U^2}{(1-\Phi_{jk}^2/U^2)(1-\Phi_{ki}^2/U^2)(1-\Phi_{\ell i}^2/U^2)}  \notag \\
& -4\frac{1-\Phi_{k\ell }\Phi_{\ell j}/U^2 - \Phi_{\ell i }\Phi_{k\ell }/U^2 + \Phi_{\ell j}\Phi_{\ell i }/U^2}{(1-\Phi_{k\ell }^2/U^2)(1-\Phi_{\ell j}^2/U^2)(1-\Phi_{\ell i }^2/U^2)} + 4\frac{1-\Phi_{\ell i }\Phi_{i k}/U^2 - \Phi_{i j }\Phi_{\ell i }/U^2 + \Phi_{i k}\Phi_{i j }/U^2}{(1-\Phi_{\ell i }^2/U^2)(1-\Phi_{i k}^2/U^2)(1-\Phi_{i j }^2/U^2)}  \notag \\
& + 3\frac{1+\Phi_{ij}\Phi_{k\ell}/U^2 - \Phi_{ij}\Phi_{\ell i}/U^2 - \Phi_{k\ell}\Phi_{\ell i}/U^2}{(1-\Phi_{ij}^2/U^2)(1-\Phi_{k\ell}^2/U^2)(1-\Phi_{\ell i}^2/U^2)} - \frac{1 +\Phi_{ij}\Phi_{k\ell}/U^2 - \Phi_{ij}\Phi_{jk}/U^2 - \Phi_{k\ell}\Phi_{jk}/U^2}{(1-\Phi_{ij}^2/U^2)(1-\Phi_{k\ell}^2/U^2)(1-\Phi_{jk}^2/U^2)}  \notag \\
&+3\frac{1-\Phi_{ij}\Phi_{jk}/U^2 - \Phi_{ij}\Phi_{\ell i}/U^2 + \Phi_{jk}\Phi_{\ell i}/U^2}{(1-\Phi_{ij}^2/U^2)(1-\Phi_{jk}^2/U^2)(1-\Phi_{\ell i}^2/U^2)} - \frac{1 - \Phi_{k\ell}\Phi_{jk}/U^2 - \Phi_{k\ell}\Phi_{\ell i}/U^2 + \Phi_{jk}\Phi_{\ell i}/U^2}{(1-\Phi_{k\ell}^2/U^2)(1-\Phi_{jk}^2/U^2)(1-\Phi_{\ell i}^2/U^2)}  \notag \\
&+ 3\frac{1 - \Phi_{jk}\Phi_{k\ell}/U^2 - \Phi_{jk}\Phi_{ij}/U^2 + \Phi_{k\ell}\Phi_{ij}/U^2}{(1-\Phi_{jk}^2/U^2)(1-\Phi_{k\ell}^2/U^2)(1-\Phi_{ij}^2/U^2)} - \frac{1 - \Phi_{\ell i}\Phi_{k\ell}/U^2 - \Phi_{\ell i}\Phi_{ij}/U^2 + \Phi_{k\ell}\Phi_{ij}/U^2}{(1-\Phi_{\ell i}^2/U^2)(1-\Phi_{k\ell}^2/U^2)(1-\Phi_{ij}^2/U^2)}  \notag \\
& \left. + 3\frac{1-\Phi_{k\ell}\Phi_{\ell i}/U^2 - \Phi_{k\ell}\Phi_{jk}/U^2 + \Phi_{\ell i}\Phi_{jk}/U^2}{(1-\Phi_{k\ell}^2/U^2)(1-\Phi_{\ell i}^2/U^2)(1-\Phi_{jk}^2/U^2)} - \frac{1-\Phi_{ij}\Phi_{\ell i}/U^2 - \Phi_{ij}\Phi_{jk}/U^2 + \Phi_{\ell i}\Phi_{jk}/U^2}{(1-\Phi_{ij}^2/U^2)(1-\Phi_{\ell i}^2/U^2)(1-\Phi_{jk}^2/U^2)} \right\}. \notag
\end{align}

Finally, putting in the potential energy differences $\Phi_{ij}$ for the different sites, we can find the specific form for the triangular lattice. We define $\theta_1 = \theta$, $\theta_2 = \theta-\pi/3$, $\theta_3 = \theta - 2\pi/3$, $\theta_1' = \theta-\pi/6$, $\theta_2' = \theta - \pi/2$, and $\theta_3' = \theta - 5\pi/6$. Then, the coupling constants for $J^{(1)}_1$, the nearest-neighbor along  $\mathbf{a}_1$, for $J^{(1)}_2$, the next nearest-neighbor along $\mathbf{a}_1'$, $J^{(1)}_3$, the next next neighbor neighbor along $\mathbf{a}_1''$, and $J^{(1)}_r$, the ring exchange coupling on $R_1$, are given by the following expressions. The nearest-neighbor coupling is the most complicated because there are the most number of virtual hopping processes possible for this coupling. When an electron undergoes a virtual hopping process, its potential energy changes according to the electric field component along that direction. This is the reason why numerous angles appear in any given coupling, not just the angle along which the coupling occurs. 
\begin{align}
J_1^{(1)}(\theta) ={}& \frac{4t^2}{U}\frac{1}{1-\alpha^2\cos^2\theta_1} + \frac{t^4}{U^3}\left[-\frac{24(1+\alpha^2\cos^2\theta_1)}{(1-\alpha^2\cos^2\theta_1)^3} - \frac{16\left(1 - \alpha^4 \cos^2\theta_1 \cos^2\theta_3\right)}{\left(1 - \alpha^2 \cos^2\theta_1\right)^{2} \left(1 - \alpha^2 \cos^2\theta_3\right)^{2}} \right. \label{eq:j1a} \\ 
& - \frac{16\left(1 - \alpha^4 \cos^2\theta_1 \cos^2\theta_2\right)}{\left(1 - \alpha^2 \cos^2\theta_1\right)^{2} \left(1 - \alpha^2 \cos^2\theta_2\right)^{2}} + \frac{4 \left(\sqrt{3} \alpha^2 \cos\theta_1 \cos\theta_1' + \alpha^2 \cos\theta_1 \cos\theta_2 + \sqrt{3} \alpha^2 \cos\theta_1' \cos\theta_2 + 1\right)}{\left(1 - \alpha^2 \cos^2\theta_1\right) \left(1 - 3 \alpha^2 \cos^2\theta_1'\right) \left(1 - \alpha^2 \cos^2\theta_2\right)} \nonumber \\ 
 & + \frac{4 \left(\alpha^2 \cos^2\theta_1 + 2 \sqrt{3} \alpha^2 \cos\theta_1 \cos\theta_1' + 1\right)}{\left(1 - \alpha^2 \cos^2\theta_1\right)^{2} \left(1 - 3 \alpha^2 \cos^2\theta_1'\right)} + \frac{8 \left(5 \alpha^2 \cos^2\theta_1 + 1\right)}{\left(1 - 4 \alpha^2 \cos^2\theta_1\right) \left(1 - \alpha^2 \cos^2\theta_1\right)^{2}} \nonumber \\ 
&+ \frac{4 \left(- \alpha^2 \cos\theta_1 \cos\theta_3 - \sqrt{3} \alpha^2 \cos\theta_1 \cos\theta_3' + \sqrt{3} \alpha^2 \cos\theta_3 \cos\theta_3' + 1\right)}{\left(1 - \alpha^2 \cos^2\theta_1\right) \left(1 - \alpha^2 \cos^2\theta_3\right) \left(1 - 3 \alpha^2 \cos^2\theta_3'\right)}  \nonumber \\ 
&+ \frac{4 \left(\alpha^2 \cos^2\theta_1 - 2 \alpha^2 \cos\theta_1 \cos\theta_3 + 1\right)}{\left(1 - \alpha^2 \cos^2\theta_1\right)^{2} \left(1 - \alpha^2 \cos^2\theta_3\right)} + \frac{16\left(1 - \alpha^4 \cos^2\theta_2 \cos^2\theta_3\right)}{\left(1 - \alpha^2 \cos^2\theta_2\right)^{2} \left(1 - \alpha^2 \cos^2\theta_3\right)^{2}} \nonumber \\ 
&+ \frac{4(\alpha^2 \cos^2\theta_2 + 2\alpha^2 \cos\theta_2 \cos\theta_3 + 1)}{\left(1 - \alpha^2 \cos^2\theta_2\right)^{2} \left(1 - \alpha^2 \cos^2\theta_3\right)} + \frac{4(\sqrt{3} \alpha^2 \cos\theta_1' \cos\theta_2 + \alpha^2 \cos^2\theta_2 + 1)}{\left(1 - 3 \alpha^2 \cos^2\theta_1'\right) \left(1 - \alpha^2 \cos^2\theta_2\right)^{2}} \nonumber \\
& - \frac{8}{\left(1 - \alpha^2 \cos^2\theta_1\right) \left(1 - \alpha^2 \cos^2\theta_2\right)} - \frac{4(\alpha^2 \cos^2\theta_1 - 2 \alpha^2 \cos\theta_1 \cos\theta_3 + 1)}{\left(1 - \alpha^2 \cos^2\theta_1\right)^{2} \left(1 - \alpha^2 \cos^2\theta_3\right)} \nonumber \\
&- \frac{4(\alpha^2 \cos^2\theta_1 + 2\sqrt{3} \alpha^2 \cos\theta_1 \cos\theta_1' + 1)}{\left(1 - \alpha^2 \cos^2\theta_1\right)^{2} \left(1 - 3 \alpha^2 \cos^2\theta_1'\right)} - \frac{8}{\left(1 - \alpha^2 \cos^2\theta_1\right) \left(1 - \alpha^2 \cos^2\theta_3\right)} \nonumber \\
&+ \frac{4( \alpha^2 \cos^2\theta_3 + 2 \sqrt{3} \alpha^2 \cos\theta_3 \cos\theta_3' + 1)}{\left(1 - \alpha^2 \cos^2\theta_3\right)^{2} \left(1 - 3 \alpha^2 \cos^2\theta_3'\right)} + \frac{4(2\alpha^2 \cos\theta_2 \cos\theta_3 + \alpha^2 \cos^2\theta_3 + 1)}{\left(1 - \alpha^2 \cos^2\theta_2\right) \left(1 - \alpha^2 \cos^2\theta_3\right)^{2}} \nonumber \\ 
&+ \frac{4}{\left(1 - \alpha^2 \cos^2\theta_2\right) \left(1 - \alpha^2 \cos^2\theta_3\right)} - \frac{2(2\sqrt{3} \alpha^2 \cos\theta_2' \cos\theta_3 + \alpha^2 \cos^2\theta_3 + 1)}{\left(1 - 3 \alpha^2 \cos^2\theta_2'\right) \left(1 - \alpha^2 \cos^2\theta_3\right)^{2}} + \nonumber \\
& - \frac{4(\sqrt{3} \alpha^2 \cos\theta_2 \cos\theta_2' + \alpha^2 \cos\theta_2 \cos\theta_3 + \sqrt{3} \alpha^2 \cos\theta_2' \cos\theta_3 + 1)}{\left(1 - \alpha^2 \cos^2\theta_2\right) \left(1 - 3 \alpha^2 \cos^2\theta_2'\right) \left(1 - \alpha^2 \cos^2\theta_3\right)} \nonumber \\
&- \frac{2(\alpha^2 \cos^2\theta_2 + 2\sqrt{3} \alpha^2 \cos\theta_2 \cos\theta_2' + 1)}{\left(1 - \alpha^2 \cos^2\theta_2\right)^{2} \left(1 - 3 \alpha^2 \cos^2\theta_2'\right)} - \frac{2(- 2 \alpha^2 \cos\theta_1 \cos\theta_3 + \alpha^2 \cos^2\theta_3 + 1)}{\left(1 - \alpha^2 \cos^2\theta_1\right) \left(1 - \alpha^2 \cos^2\theta_3\right)^{2}} \nonumber \\
& \left. - \frac{2( \alpha^2 \cos\theta_1 \cos\theta_2 +  \alpha^2 \cos^2\theta_2 + 1)}{\left(1 - \alpha^2 \cos^2\theta_1\right) \left(1 - \alpha^2 \cos^2\theta_2\right)^{2}} +  \frac{4 \left(-\alpha^2 \cos\theta_1 \cos\theta_2 + \alpha^2 \cos\theta_1 \cos\theta_3 + 3\alpha^2 \cos\theta_2 \cos\theta_3 + 1\right)}{\left(1 - \alpha^2 \cos^2\theta_1\right) \left(1 - \alpha^2 \cos^2\theta_2\right) \left(1 - \alpha^2 \cos^2\theta_3\right)} \right] \nonumber \\
J_2^{(1)}(\theta) ={}& \frac{8t^4}{U^3}\left[\frac{2\left(1 - \alpha^4 \cos^2\theta_1 \cos^2\theta_2\right)}{\left(1 - \alpha^2 \cos^2\theta_1\right)^{2} \left(1 - \alpha^2 \cos^2\theta_2\right)^{2}} - \frac{\sqrt{3} \alpha^2 \cos\theta_1 \cos\theta_1' + \alpha^2 \cos\theta_1 \cos\theta_2 + \sqrt{3} \alpha^2 \cos\theta_1' \cos\theta_2 + 1}{\left(1 - \alpha^2 \cos^2\theta_1\right) \left(1 - 3 \alpha^2 \cos^2\theta_1'\right) \left(1 - \alpha^2 \cos^2\theta_2\right)} \right] \label{eq:j2ap} \\
&+\frac{2t^4}{U^3}\left[ \frac{2}{\left(1 - \alpha^2 \cos^2\theta_1\right) \left(1 - \alpha^2 \cos^2\theta_2\right)} - \frac{\alpha^2 \cos^2\theta_2 + 2\alpha^2 \cos\theta_2 \cos\theta_3 +1}{\left(1 - \alpha^2 \cos^2\theta_2\right)^{2} \left(1 - \alpha^2 \cos^2\theta_3\right)} \right. \nonumber \\
&- \frac{2\sqrt{3} \alpha^2 \cos\theta_1' \cos\theta_2 + \alpha^2 \cos^2\theta_2 + 1}{\left(1 - 3 \alpha^2 \cos^2\theta_1'\right) \left(1 - \alpha^2 \cos^2\theta_2\right)^{2}} \nonumber \\
& + \frac{2(\alpha^2 \cos\theta_1 \cos\theta_2 + \alpha^2 \cos\theta_1 \cos\theta_3 - \alpha^2 \cos\theta_2 \cos\theta_3 - 1)}{\left(1 - \alpha^2 \cos^2\theta_1\right) \left(1 - \alpha^2 \cos^2\theta_2\right) \left(1 - \alpha^2 \cos^2\theta_3\right)} \nonumber \\
&+ \frac{2 (\sqrt{3} \alpha^2 \cos\theta_1 \cos\theta_1' + \alpha^2 \cos\theta_1 \cos\theta_2 + \sqrt{3} \alpha^2 \cos\theta_1' \cos\theta_2 + 1)}{\left(1 - \alpha^2 \cos^2\theta_1\right) \left(1 - 3 \alpha^2 \cos^2\theta_1'\right) \left(1 - \alpha^2 \cos^2\theta_2\right)} \nonumber \\
&\left. - \frac{ \alpha^2 \cos^2\theta_1 - 2 \alpha^2 \cos\theta_1 \cos\theta_3 +1}{\left(1 - \alpha^2 \cos^2\theta_1\right)^{2} \left(1 - \alpha^2 \cos^2\theta_3\right)} - \frac{\alpha^2 \cos^2\theta_1 + 2 \sqrt{3} \alpha^2 \cos\theta_1 \cos\theta_1' + 1}{\left(1 - \alpha^2 \cos^2\theta_1\right)^{2} \left(1 - 3 \alpha^2 \cos^2\theta_1'\right)}\right] \nonumber \\
J_3^{{(1)}}(\theta) ={}& \frac{4t^4}{U^3} \left[\frac{2\left(1 + \alpha^2 \cos^2\theta_1\right)}{\left(1 - \alpha^2 \cos^2\theta_1\right)^{3}} - \frac{5 \alpha^2 \cos^2\theta_1 + 1}{\left(1 - 4 \alpha^2 \cos^2\theta_1\right) \left(1 - \alpha^2 \cos^2\theta_1\right)^{2}}\right] \label{eq:j3a} \\
J_r^{(1)}(\theta) ={}& \frac{8t^4}{U^3}\left[ \frac{2}{\left(1 - \alpha^2 \cos^2\theta_1\right) \left(1 - \alpha^2 \cos^2\theta_2\right)} + \frac{\alpha^2 \cos^2\theta_2 + 2\alpha^2 \cos\theta_2 \cos\theta_3 + 1}{\left(1 - \alpha^2 \cos^2\theta_2\right)^{2} \left(1 - \alpha^2 \cos^2\theta_3\right)} + \frac{\alpha^2 \cos^2\theta_1 - 2\alpha^2 \cos\theta_1 \cos\theta_3 + 1}{\left(1 - \alpha^2 \cos^2\theta_1\right)^{2} \left(1 - \alpha^2 \cos^2\theta_3\right)} \right. \label{eq:jrap} \\
&+ \frac{2\left(- \alpha^2 \cos\theta_1 \cos\theta_2 - \alpha^2 \cos\theta_1 \cos\theta_3 + \alpha^2 \cos\theta_2 \cos\theta_3 + 1\right)}{\left(1 - \alpha^2 \cos^2\theta_1\right) \left(1 - \alpha^2 \cos^2\theta_2\right) \left(1 - \alpha^2 \cos^2\theta_3\right)} \nonumber \\
&+ \frac{2\left(\sqrt{3} \alpha^2 \cos\theta_1 \cos\theta_1' + \alpha^2 \cos\theta_1 \cos\theta_2 + \sqrt{3} \alpha^2 \cos\theta_1' \cos\theta_2 + 1\right)}{\left(1 - \alpha^2 \cos^2\theta_1\right) \left(1 - 3 \alpha^2 \cos^2\theta_1'\right) \left(1 - \alpha^2 \cos^2\theta_2\right)} \nonumber \\
&\left. + \frac{2\sqrt{3} \alpha^2 \cos\theta_1' \cos\theta_2 + \alpha^2 \cos^2\theta_2 + 1}{\left(1 - 3 \alpha^2 \cos^2\theta_1'\right) \left(1 - \alpha^2 \cos^2\theta_2\right)^{2}} + \frac{\alpha^2 \cos^2\theta_1 + 2 \sqrt{3} \alpha^2 \cos\theta_1 \cos\theta_1' + 1}{\left(1 - \alpha^2 \cos^2\theta_1\right)^{2} \left(1 - 3 \alpha^2 \cos^2\theta_1'\right)} \right]. \nonumber 
\end{align}

We note that, in order to obtain the coupling constants along different bonds, we may use $J^{(2)}_1(\theta) = J^{(1)}_1(\theta - \pi/3)$, and $J^{(3)}_1(\theta) = J^{(1)}_1(\theta - 2\pi/3)$. The same relations hold for the second and third nearest neighbors, as well as the ring exchange.
\end{widetext}

\section{DMRG Calculation} \label{app:dmrg}
We use the TenPy library\cite{hauschild_efficient_2018} for the iDMRG calculations. For any given point in the phase diagram we do the DMRG in 3 steps.
\begin{enumerate}
    \item First, initialize the system with an up/down product state, and maximum bond dimension of $b = 20$. Perform this initialization run with 5 sweeps and a chiral symmetry breaking term of $J_\chi = 10^{-5}$. 
    \item Now, set $J_\chi = 0$. Take the output of the first step as the initial state for a run with the density matrix mixer on to escape local minima, with a maximum of 40 sweeps and bond dimension of $b = 1600$.
    \item Take the output of the second step as the initial state for a run with the density matrix mixer off because we assume that we are in the global minimum basin, and run until the energy converges to $\Delta E = 10^{-8}$. This is also done with bond dimension $b = 1600$. 
\end{enumerate}

Sometimes, the simulation converges in energy, but does not converge to the true ground state. This is more likely when the ring exchange is large because there are an increased number of competing states. For this reason, after determining which phases appear in the phase diagram, we run points in the phase diagram again, but with specific initial conditions corresponding to the different phases of model. For example, for $E/U = 0$ and $t/U$ close to 0.1, we would use both the CSL and VBS initial conditions for such a point in parameter space, and then check which resulting wave function has a lower energy. Selecting the one which has the lowest energy yields the result that we report. 

For all calculations, we perform the simulations on a cylinder whose circumference is $L_y = 6$ sites, and which is infinite in the $\mathbf{a}_1$ direction. The unit cell we use is $L_x=2$ sites long in the $\mathbf{a}_1$ direction. We compute spin and dimer correlations to 24 unit cells along the direction of the cylinder axis. 

\section{Valence Bond Solid Phases}
In this section, we will provide some more details of the valence bond solids described in the main text. 

\subsection{Dimer Coverings of the Triangular lattice}
\begin{figure}
    \centering
    \includegraphics{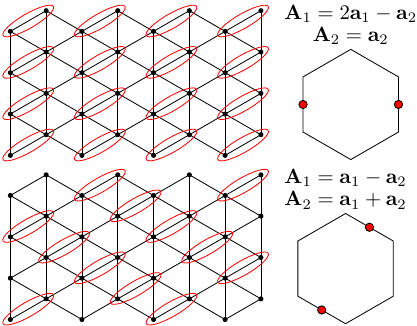}
    \caption{The two different coverings of the triangular lattice by dimers along the $\mathbf{a}_1$ bond. The two translation vectors of the unit cell are given for each covering, which allows us to find the peaks of each structure factor $D_1(\bm{k})$, given on the right.}
    \label{fig:dimers_a}
\end{figure}

\begin{figure}
    \centering
    \includegraphics{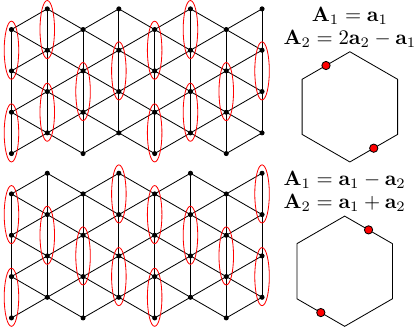}
    \caption{The two different coverings of the triangular lattice by dimers along the $\mathbf{a}_2$ bond. The two translation vectors of the unit cell are given for each covering, which allows us to find the peaks of each structure factor $D_2(\bm{k})$, given on the right.}
    \label{fig:dimers_b}
\end{figure}

\begin{figure}
    \centering
    \includegraphics{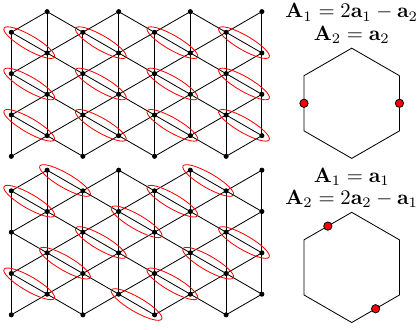}
    \caption{The two different coverings of the triangular lattice by dimers along the $\mathbf{a}_3$ bond. The two translation vectors of the unit cell are given for each covering, which allows us to find the peaks of each structure factor $D_3(\bm{k})$, given on the right.}
    \label{fig:dimers_c}
\end{figure}

As mentioned in the main text, there are 6 simple dimer coverings of the triangular lattice. This is not an exhaustive list of all coverings, but these 6 simple ones are sufficient to yield consistent results with the dimer structure factors for VBS${}_1$ and VBS${}_2$. For each covering, we list the translation vectors of the unit cell in terms of the primitive lattice vectors of the triangular lattice. Furthermore, we indicate the peak locations corresponding to the unit cell translation vectors in a schematic Brillouin zone for each covering. These coverings are in Figs.~\ref{fig:dimers_a}-\ref{fig:dimers_c}.

\subsection{Valence bond solid 1} \label{app:dimer_coverings}
There are 6 possible simple dimer coverings of the triangular lattice. To identify the VBS${}_1$ state with particular dimer coverings, we compute the three different dimer structure factors. Using the peak heights of each structure factor, we can identify which are the most dominant dimer coverings. For VBS${}_1$, the highest peaks are for dimers along $\mathbf{a}_1$ and $\mathbf{a}_3$. Then, since $D_1(\mathbf{k})$ has peaks at $M'$, and $D_3(\mathbf{k})$ has peaks at $M''$, we have that the wave function is (approximately) the equal weight superposition given in Eq.~\eqref{eq:VBS1_WF}. We note that this is only an approximate wave function because other valence bond configurations are a part of the wave function, but are subdominant. 

\subsection{Valence bond solid 2}
For VBS${}_2$, the dimer structure factor with the largest peak height is the one along the $\mathbf{a}_2$. The peaks are at the $M''$ points. Identifying this with the configurations in Fig.~\ref{fig:dimers_b}, we find the wave function in Eq.~\eqref{eq:VBS2_WF}.

\subsection{Valence bond solid 3}
\begin{figure*}
\centering
\includegraphics[scale=1]{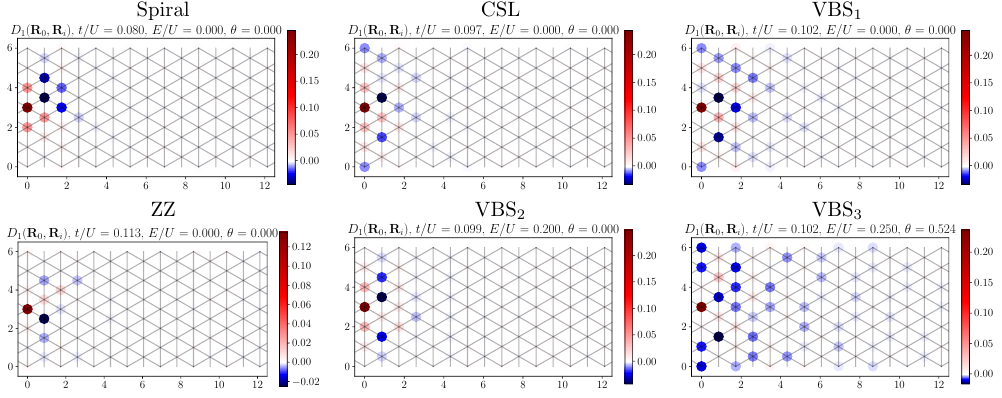}
\caption{Real space dimer structure factors for dimers along the $\mathbf{a}_1$ direction for all the phases in our model. These are defined as $D_1(\mathbf{R}_0,\mathbf{R}_i) = \langle D^1_0 D^1_i \rangle - \langle D^1_0 \rangle \langle D^1_i \rangle$, where $\mathbf{R}_0 = 3\mathbf{a}_2$ is the site halfway up the cylinder in the first ring. Note that the top row of sites is simply the bottom row copied up, due to the periodic boundary conditions.}
\label{fig:sdsfa_real_all}
\end{figure*}

\begin{figure*}
\centering
\includegraphics[scale=1]{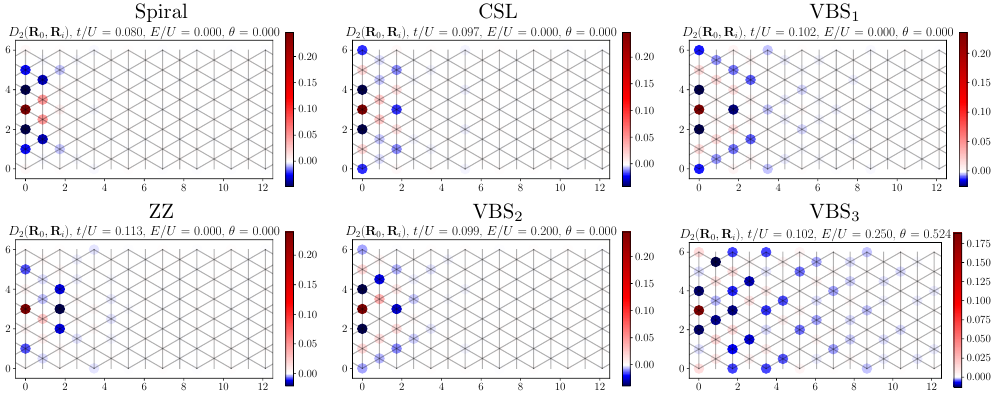}
\caption{Real space dimer structure factors for dimers along the $\mathbf{a}_2$ direction for all the phases in our model. These are defined as $D_2(\mathbf{R}_0,\mathbf{R}_i) = \langle D^2_0 D^2_i \rangle - \langle D^2_0 \rangle \langle D^2_i \rangle$, where $\mathbf{R}_0 = 3\mathbf{a}_2$ is the site halfway up the cylinder in the first ring. Note that the top row of sites is simply the bottom row copied up, due to the periodic boundary conditions.}
\label{fig:sdsfb_real_all}
\end{figure*}

\begin{figure*}
\centering
\includegraphics[scale=1]{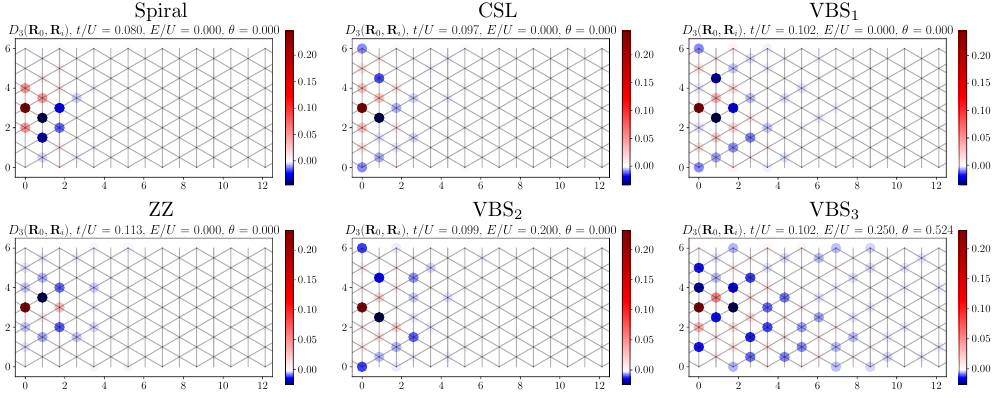}
\caption{Real space dimer structure factors for dimers along the $\mathbf{a}_3$ direction for all the phases in our model. These are defined as $D_3(\mathbf{R}_0,\mathbf{R}_i) = \langle D^3_0 D^3_i \rangle - \langle D^3_0 \rangle \langle D^3_i \rangle$, where $\mathbf{R}_0 = 3\mathbf{a}_2$ is the site halfway up the cylinder in the first ring. Note that the top row of sites is simply the bottom row copied up, due to the periodic boundary conditions.}
\label{fig:sdsfc_real_all}
\end{figure*}

The third valence bond solid has strong dimer correlations, but not at high symmetry points. The dimer correlations are long-ranged, as can be seen in Figs.~\ref{fig:sdsfa_real_all}-\ref{fig:sdsfc_real_all}. This suggests that this phase is a valence bond solid whose unit cell is larger than the size of our finite sized simulation.

\section{DMRG Data\label{app:dmrg_data}}

\subsection{Spin Structure Factor}
The spin structure factor is defined by 

\begin{equation}
S(\bm{k}) = \sum_{ij} e^{i\bm{k}\cdot(\mathbf{R}_i - \mathbf{R}_j)} (\langle \mathbf{S}_i\cdot \mathbf{S}_j \rangle - \langle \mathbf{S}_i \rangle \cdot \langle \mathbf{S}_j \rangle).
\end{equation}

The spin structure factor is primarily used to identify magnetically ordered phases. The peaks of the momentum space spin structure factor should be at high symmetry points in the Brillouin zone, and also be sharp, indicating long ranged correlations. We also include the real space spin correlations in order to demonstrate the slow decay. 

\begin{figure*}
\centering
\includegraphics[scale=1]{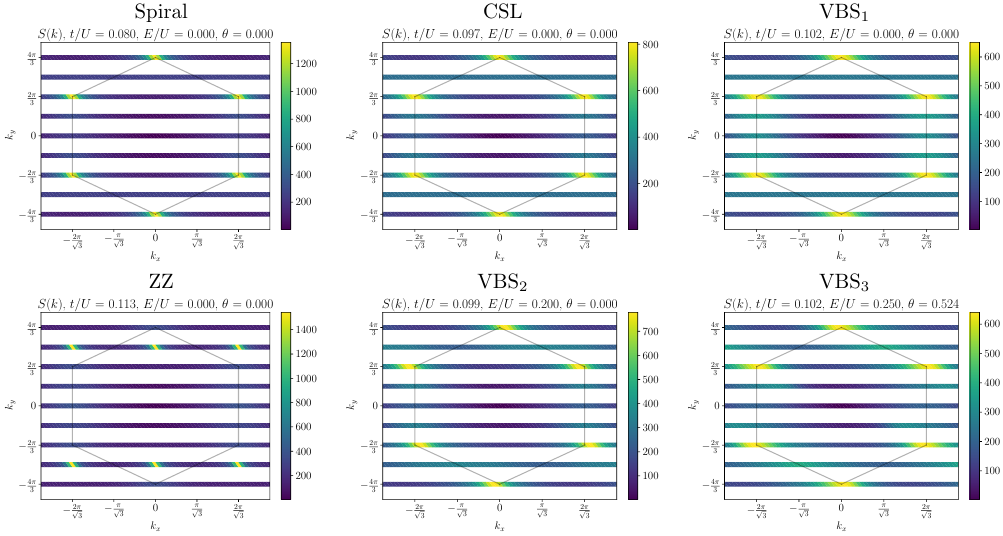}
\caption{Momentum space spin structure factor for the difference phases observed in the model.}
\label{fig:sssf_mom_all}
\end{figure*}

\begin{figure*}
\centering
\includegraphics[scale=1]{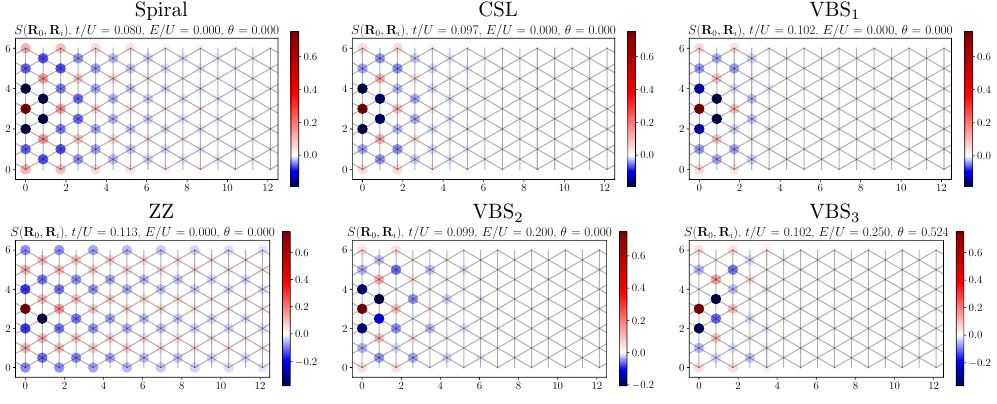}
\caption{Real space spin structure factor for the difference phases observed in the model. This is defined as $S(\mathbf{R}_0,\mathbf{R}_i) = \langle \mathbf{S}_0 \cdot \mathbf{S}_i \rangle - \langle \mathbf{S}_0 \rangle \cdot \langle \mathbf{S}_i \rangle$, where $\mathbf{R}_0 = 3\mathbf{a}_2$ is the site halfway up the cylinder in the first ring. Note that the top row of sites is simply the bottom row copied up, due to the periodic boundary conditions.}
\label{fig:sssf_real_all}
\end{figure*}

\subsection{Dimer Structure Factors}

We define the dimer operator $D^n_i = \mathbf{S}_i \cdot \mathbf{S}_{i+\mathbf{a}_n}$. This structure factor probes the formation of singlets between nearest-neighbor spins in the lattice along the $\mathbf{a}_n$ direction. Then, we define the dimer structure factor by

\begin{equation}
D_n(\bm{k}) = \sum_{ij} e^{i\bm{k}\cdot(\mathbf{R}_i - \mathbf{R}_j)} (\langle D^n_i D^n_j \rangle - \langle D^n_i \rangle\langle D^n_j \rangle).
\end{equation}

\begin{figure*}
\centering
\includegraphics[scale=1]{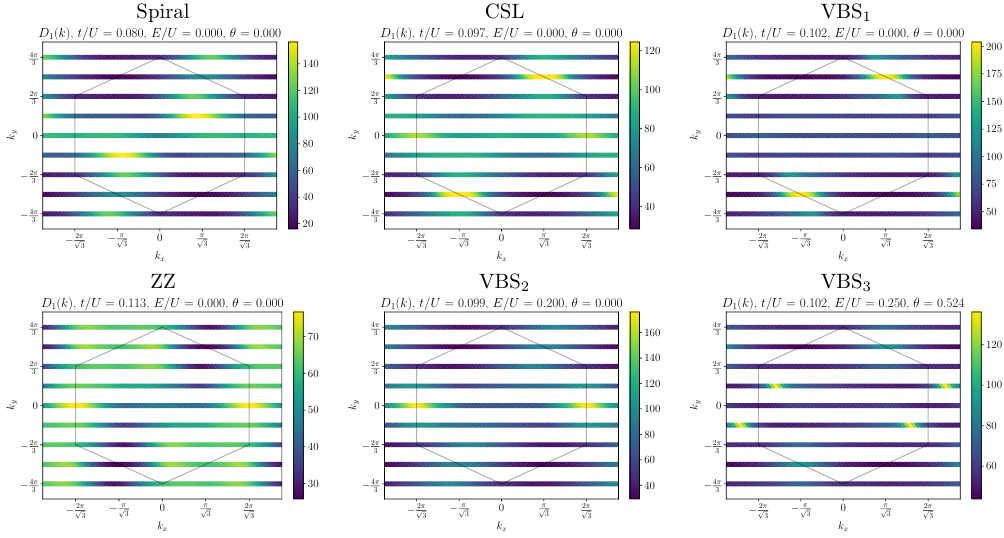}
\caption{Momentum space dimer structure factors for dimers along the $\mathbf{a}_1$ direction for all the phases in our model.}
\label{fig:sdsfa_mom_all}
\end{figure*}

\begin{figure*}
\centering
\includegraphics[scale=1]{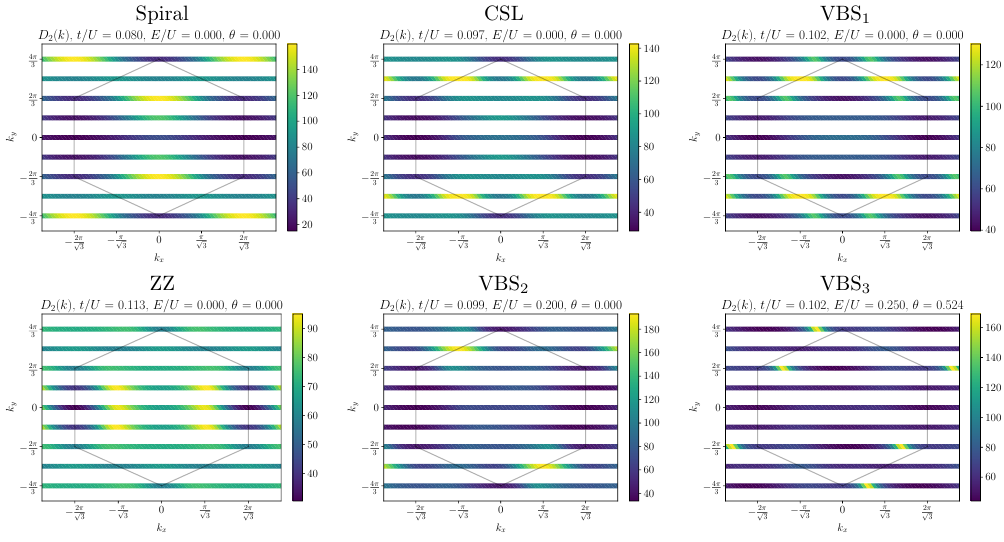}
\caption{Momentum space dimer structure factors for dimers along the $\mathbf{a}_2$ direction for all the phases in our model.}
\label{fig:sdsfb_mom_all}
\end{figure*}

\begin{figure*}
\centering
\includegraphics[scale=1]{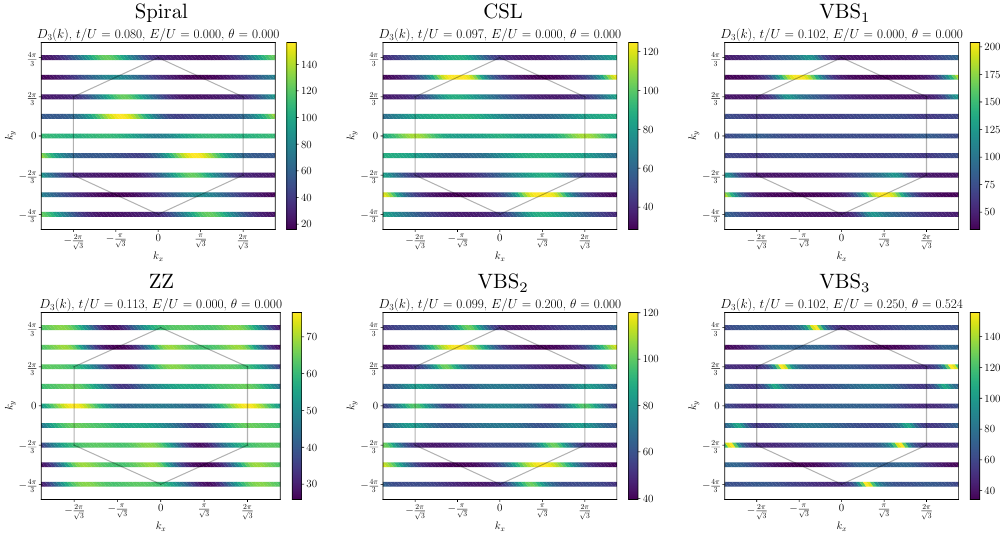}
\caption{Momentum space dimer structure factors for dimers along the $\mathbf{a}_3$ direction for all the phases in our model.}
\label{fig:sdsfc_mom_all}
\end{figure*}

\subsection{Entanglement Spectrum}
\begin{figure*}
\centering
\includegraphics[scale=1]{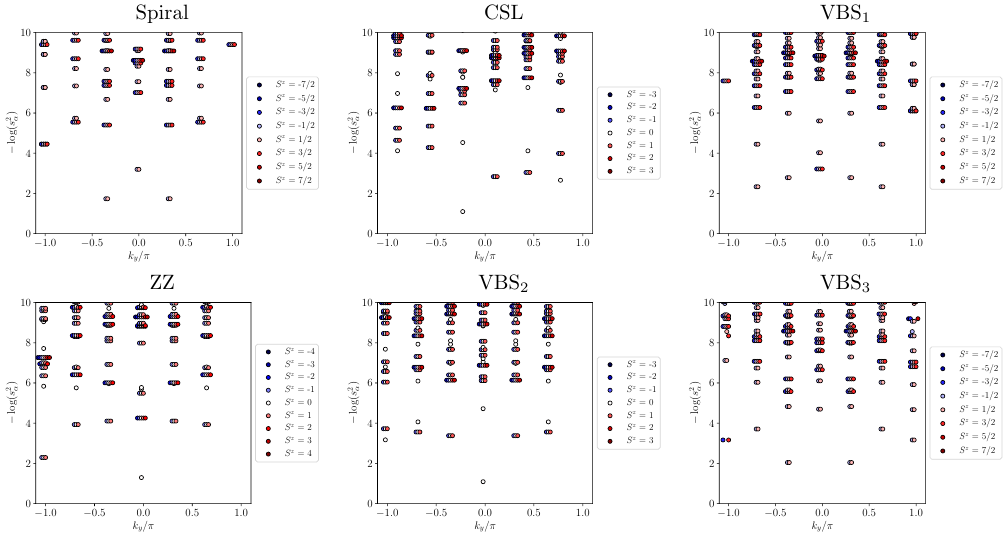}
\caption{Entanglement spectrum $-\log(s_\alpha^2)$ by transverse momentum $k_y$ for all phases in the model.}
\label{fig:es_mom}
\end{figure*}

\begin{figure*}
\centering
\includegraphics[scale=1]{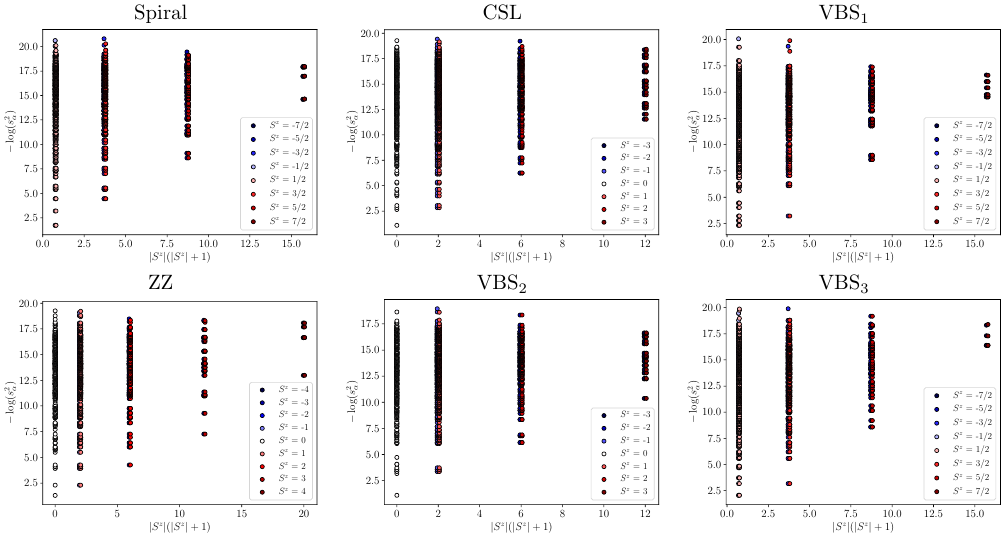}
\caption{Entanglement spectrum $-\log(s_\alpha^2)$ by $S^z$ quantum number for all phases in the model.}
\label{fig:es_charge}
\end{figure*}

As mentioned in the main text, the entanglement spectrum is obtained by considering the mixed transfer matrix $\mathcal{T}^T_{\alpha_1\alpha_2\alpha_3\alpha_4}$, which is constructed by translating the wavefunction $\ket{\psi}$ by one lattice spacing along the circumferential direction, and then contracting its physical indices with those of the original (untranslated) wavefunction. The resulting 4-index object (each $\alpha$ index running from 1 to bond dimension $b$) can be diagonalized, with dominant eigenvalue 1 if the wave function is translationally invariant (if the wave function is not translationally invariant, then the momentum is not well defined). The eigenvector $V_{\alpha_1\alpha_2}$ corresponding to eigenvalue 1 is itself diagonalizable, and its eigenvalues are $\mu_\alpha = s_\alpha^2 e^{ik_\alpha}$, where $s_\alpha$ are the Schmidt values, and $k_\alpha$ is the momentum around the cylinder~\cite{hauschild_efficient_2018, gohlke_quantum_2018}. The entanglement spectrum is therefore $-\log(s_\alpha^2)$. The corresponding figures of the momentum resolved, and spin number resolved entanglement spectra are in Figs.~\ref{fig:es_mom} and \ref{fig:es_charge}.

\bibliographystyle{apsrev4-1}

%

\end{document}